\documentclass[a4paper, onecolumn, 11pt,accepted=2021-12-14]{quantumarticle}
\pdfoutput=1

\usepackage{amsthm}
\usepackage{amssymb}
\usepackage{url}
\usepackage{mathtools}

\usepackage{float}
\usepackage{bm}
\usepackage{amsmath}
\usepackage{graphicx}
\usepackage{mleftright}
\usepackage[braket,qm]{qcircuit}
\usepackage[normalem]{ulem}

\usepackage{tikz}
\usetikzlibrary{matrix,positioning}
\tikzset{bullet/.style={circle,fill,inner sep=2pt}}

\usepackage{algpseudocode} 
\usepackage{algorithm}

\renewcommand{\H}{\mathcal{H}}

\newcommand{\diag}[1]{\text{diag}\left(#1\right)}

\usepackage[sorting=none,style=numeric,citestyle=numeric-comp]{biblatex}
\begin{document}

\title{The Efficient Preparation of Normal Distributions in Quantum Registers}

\author{Arthur G. Rattew}
\affiliation{Future Lab for Applied Research and Engineering, JPMorgan Chase \& Co.}
\author{Yue Sun}
\affiliation{Future Lab for Applied Research and Engineering, JPMorgan Chase \& Co.}
\author{Pierre Minssen}
\affiliation{Future Lab for Applied Research and Engineering, JPMorgan Chase \& Co.}
\author{Marco Pistoia}
\affiliation{Future Lab for Applied Research and Engineering, JPMorgan Chase \& Co.}

\maketitle

\begin{abstract}
The efficient preparation of input distributions is an important problem in obtaining quantum advantage in a wide range of domains.
We propose a novel quantum algorithm for the efficient preparation of arbitrary normal distributions in quantum registers.
To the best of our knowledge, our work is the first to leverage the power of Mid-Circuit Measurement and Reuse (MCMR), in a way that is broadly applicable to a range of state-preparation problems. Specifically, our algorithm employs a repeat-until-success scheme, and only requires a constant-bounded number of repetitions in expectation.
In the experiments presented, the use of MCMR enables up to a $862.6\times$ reduction in required qubits. 
Furthermore, the algorithm is provably resistant to both phase-flip and bit-flip errors, leading to a first-of-its-kind empirical demonstration on real quantum hardware, the MCMR-enabled Honeywell System Models H0 and H1-2. 
\end{abstract}

\section{Introduction}
The efficient preparation of input distributions is particularly important for a wide range of quantum algorithms, such as those for amplitude estimation~\cite{vazquez2020efficient}, option pricing~\cite{stamatopoulos2020option}, principal-component analysis~\cite{lloyd2014quantum}, matrix inversion~\cite{harrow2009quantum}, and machine learning~\cite{ciliberto2018quantum, mitarai2018quantum, rebentrost2014quantum}, which all offer the potential for quantum advantage, notably in financial applications \cite{pistoia2021quantum}, so long as their initial distributions may be generated without introducing computational bottlenecks.

Constructing an arbitrary quantum state necessitates exponential-depth circuits~\cite{plesch2011quantum}. As a result, any efficient state preparation technique must either be approximate in nature, or exploit information specific to the distribution being generated. Moreover, the quantum computers of the foreseeable future belong to the class of Noisy Intermediate-Scale Quantum (NISQ) hardware, characterized by small qubit counts, limited two-qubit gate fidelity, and short coherence times~\cite{preskill2018quantum}. As a result, for an algorithm to offer quantum advantage in the near future, it is even more important that state-generation procedures use as shallow circuits with as few ancillary qubits as possible, and produce high-fidelity states even in the presence of low gate-execution fidelity.

Initially proposed by Lloyd and Weedbrook in 2018 for the purpose of generating quantum states, quantum generative adversarial networks (QGANs) employ two agents: a generator and a discriminator~\cite{lloyd2014quantum}. The generator is tasked with producing the desired distribution, which is then evaluated by the discriminator~\cite{pan2019recent}. Numerous papers have since built upon this work, demonstrating QGANs on real quantum processors, and extending QGANs for both classical sampling as well as for loading coherent quantum states~\cite{zoufal2019quantum, hu2019quantum}. However, these techniques are approximate, and are not necessary in cases where efficient circuits may be theoretically derived.

In the creation of exact distributions, or those motivated analytically, a wide range of approaches have been explored. Grover and Rudolph published a procedure for generating efficiently integrable (e.g., log-concave) probability-density functions~\cite{grover2002creating}. While there are some conflicting opinions regarding the theoretical asymptotic efficiency of the described procedure~\cite{holmes2020efficient, vazquez2020efficient}), all agree that this work does not offer an approach that is efficiently realizable in practice on extant NISQ hardware.  Kitaev and Webb built upon the aforementioned work by Grover and Rudolph, describing a method of generating multivariate normal distributions through \emph{resampling}~\cite{kitaev2008wavefunction}, similar to the qubit-scaling procedure we present in this paper, but fundamentally different in a  way that causes their approach to rapidly accumulate error. 
Moreover, their resampling technique only allows for the efficient creation of Gaussian distributions with small variances, and requires the use of multiple ancillary qubits, limiting its practicality in the NISQ era. Our approach has no such limitations.
H\"{a}ner \textit{et al.} presented a number of quantum circuits with polynomial depth, implementing commonly encountered functions through piece-wise polynomial approximations~\cite{haner2018optimizing}.
Nevertheless, this approach comes with a significant overhead in both required ancillary qubits, and resulting circuit depths~\cite{vazquez2020efficient}. Moreover, we do not expect this approach to efficiently implement a range of functions, such as the exponential function (and thus normal distributions), as the piece-wise polynomial approach would require an infeasible number of pieces to create accurate models on near-term hardware.

Non-unitary transformations may enable distributions to be obtained more efficiently~\cite{ahansaz2019quantum, kendon2008optimal}.
Two common ways to implement non-unitary transformations are (1) scaling to higher dimensional spaces through the use of ancillary qubits~\cite{haner2018optimizing}), and (2) introducing non-linearities by performing partial measurements throughout the coherent execution of a circuit.
The use of an asymptotic number of ancillary qubits is not desirable for the purpose of introducing non-unitary transformations, as qubit counts are limited in the NISQ era. As such, we believe it is preferable to use Mid-Circuit Measurement and Reuse (MCMR)~\cite{Pino_2021}.
The use MCMR has been demonstrated to generate more NISQ-friendly circuits with lower resource footprints.
Specifically, Yalovetzky~\textit{et al.} leveraged MCMR and quantum conditional logic for Quantum Phase Estimation (QPE), to reduce the number of ancilla qubits and two-qubit gates, and lower the requirement on qubit connectivity \cite{yalovetzky2021nisqhhl}. On the hardware, this QPE variant produced results with higher fidelity than the orignal QPE.
In the context of this work, the MCMR-free approach requires a number of ancillas that scales asymptotically in the variance of the target distribution being generated.
As a result, in the experiments performed, the MCMR-based solution we propose uses up to $862.6\times$ fewer total qubits and $8617\times$ fewer ancillary qubits than its MCMR-free counterpart, thereby making it a more usable state-preparation procedure in the NISQ era. 

Repeat-until-success (RUS) paradigms are often used by approaches that introduce non-linearities through the use of mid-circuit partial measurements~\cite{paetznick2013repeat}. In these methodologies, ancillary \textit{flag} qubits are entangled with primary data registers, and the flags are then measured to determine if a particular operation was applied successfully. If the operation was not applied successfully, it may be possible to apply an operation conditioned on the classical measurement to then obtain the desired output. Alternatively, the circuit may simply be re-executed until the desired transformation is obtained. Guerreschi observes that RUS approaches have been used in applications such as implementing quantum neurons with non-linear activation functions~\cite{cao2017quantum, hu2018towards}, and in the synthesis of arbitrary single-qubit rotations~\cite{bocharov2015efficient}. Moreover, RUS has been applied to a number of quantum arithmetic problems~\cite{wiebe2014quantum}. Our work appears to be the first demonstration of RUS in a state-preparation algorithm. Moreover, we also demonstrate that our usage of RUS gives our algorithm an intrinsic robustness against hardware noise, further enhancing the procedure's viability in the NISQ era. 

\renewcommand{\H}{\mathcal{H}}

\section{The Algorithm}
\label{sec:algorithm}

Our algorithm prepares a quantized normal distribution in a quantum register, and receives as its input five parameters: $\hat{\mu}, \hat{\sigma}^2$, $x_0, l$ and $n$.
Here, $\hat{\mu}$ and $\hat{\sigma}^2$ specify the normal distribution $\mathcal{N}(\hat{\mu}, \hat{\sigma}^2)$,  $x_0$ and $l$ specify the interval $[x_0, x_0 + l]$ upon which the distribution is produced, and $2^{n}$ is the desired resolution. 
We assume that $[x_0, x_0 + l]$ contains the mean of the normal distribution, as well as effectively all of the probability density.
First, we introduce notation to make explicit the distinction between continuous and discrete state spaces. We allow $\{\ket{x}\}_x$ to index the set of continuous states, with $x\in [x_0, x_0 + l]$, and we allow $\{\ket{x_j}\}_j$ to index the set of discrete states with $j \in \mathbb{Z}_N$. 
We assume we have an $n$ qubit system, with $N=2^n$ states. Then, we allow $\Delta x = \frac{l}{N}$ to be the interval between our discrete states. 
The continuous and discrete states are related with,
\begin{align}
    \ket{x_j} = \ket{x_0 + j\Delta x}.
\end{align}
We define the set of integer-valued computational (standard) basis states $\{\ket{j}\}_j$ with $j\in \mathbb{Z}_N$, which are isomorphic to the discretized set of states $\{\ket{x_j}\}_j$ through the mapping $\ket{j} = \ket{(x_j - x_0)\Delta x^{-1}}$.
The algorithm aims to produce a quantum state $\ket{\psi}$ such that the amplitude of each standard basis vector $\ket{j}$ (up to a constant normalization factor) is given by,
\begin{align*}
    \ip{j}{\psi} = \ip{x_j}{\psi} = \frac{1}{\sqrt{2\pi \hat{\sigma}^2}}\int_{x_j}^{x_{j+1}}  \exp\left[{-\frac{(x - \hat{\mu})^2}{2\hat{\sigma}^2} }\right] dx.
\end{align*}
Moreover, we will assume that $\mu$ and $\sigma^2$ are in units corresponding to the integer-valued basis states, while $\hat{\mu}$ and $\hat{\sigma}^2$ are in units corresponding to the continuous input domain. As such, we have that $\hat{\mu} = \frac{l}{N}\mu$ and $\hat{\sigma}^2 = \frac{l^2}{N^2}\sigma^2$.

\subsection{Generating Normal Distributions through Discrete Random Walks}

Consider a random walk on the set of integers $\{j\}_j$ with $j\in \mathbb{Z}_N$ with $t$ steps, where in each step there is an equal probability of transitioning from $j$ to $j$ and $j+1$ \footnote{Note that this principle also applies if the transitions maps $j$ to $j-1$ or $j+1$. However, we choose the mapping $j\to j$ and $j\to j+1$ as it avoids a rectifiable problem where alternating states have zero amplitude, and moreover avoids the use of an additional ancillary qubit.}. The probability of such a walk terminating on any given integer clearly follows a binomial distribution, which tends to the normal distribution as $N\to\infty$ by the central limit theorem.
As such, an operator implementing the quantum dynamics,
\begin{align}
\label{eqn:hamiltonian_definition_original}
    \H \ket{j} = \ket{j} + \ket{j+1},
\end{align}
coherently performs such a random walk, and thus produces a quantum state with amplitudes following a binomial distribution. 
Note that the final procedure presented doesn't produce a binomial distribution, but rather directly produces a discrete normal distribution.
Therefore, we may simulate such a discrete random walk with $t$ steps by applying $\H$ $t$ times to $\ket{\psi_i}$, hence by constructing $\H^t\ket{\psi_i}$ for some starting basis state $\ket{\psi_i}$. For those familiar with Galton machines, it may be more clear to consider this random walk in the context of simulating a Galton machine, and we indeed consider this perspective in Supplementary Information (SI)~\ref{appendix:algo_intuition}, where we also provide a more comprehensive overview of the intuition behind the algorithm.


\subsubsection{Deriving \texorpdfstring{$t-\hat{\sigma}^2$}{t-sigma-hat\^2} and \texorpdfstring{$t-\hat{\mu}$}{t-mu-hat} Relationships}
\label{appendix:deriving_var_and_mean_expressions}
Applying $\H^t$ to some initial state produces a binomial distribution (on the amplitudes) equivalent to that produced by the aforementioned discrete random walk with $t$ steps. As such, the mean of the produced distribution is given by $\mu = \frac{t}{2}$, while the standard deviation is given by $\sigma = \frac{1}{2}\sqrt{t}$. Note that if we wish to obtain a given variance in the probability distribution rather than in the amplitudes, by property of a normal distribution, we may simply produce $2\times$ the desired probability variance as the amplitude variance. Throughout the remainder of this document, unless we explicitly state otherwise (e.g. by writing a probability mass function explicitly), we assume that our distributions are in terms of the amplitudes of quantum states rather than their probabilities. Then, we may compute the number of applications of $\H$ required to obtain a normal distribution with variance corresponding to $\hat{\sigma}^2$ as follows,
\begin{align*}
    \hat{\sigma}^2 = \frac{l^2}{N^2}\sigma^2 = \frac{l^2}{4N^2}t,
\end{align*}
\begin{align}\label{eqn:t_in_terms_of_var}
    \implies t = \frac{4N^2}{l^2}\hat{\sigma}^2.
\end{align}

\subsubsection{Qubit Scaling Procedure}

Equation~\ref{eqn:t_in_terms_of_var} may appear to suggest that to obtain a distribution on $n$ qubits corresponding to a variance $\hat{\sigma}^2$, a number of applications of $\H$ scaling with $O(2^{2n})$ would be required. This would be problematic, as even if each application of $\H$ could be implemented in polynomial depth (as we later show, it can) the overall circuit would necessitate exponential depth to produce a normal distribution with arbitrary variance.

Fortunately, as we show in SI~\ref{appendix:efficient_scaling_qubit_addition}, it is possible to avoid this exponential blowup. In essence, we do so by first loading a nearly exact distribution on some small constant number of qubits, $n_1$, by applying $\H$ $t_1$ times to the $n_1$ qubit $\ket{0}$ state.

We then iteratively add a new qubit in the $\ket{+}$ state (a uniform superposition on one qubit) as the least significant qubit\footnote{I.e. as qubit $n_{i+1}$, where $n_i$ refers to the number of qubits at the start of the $i^{th}$ step of the procedure. Thus, $n_{i+1} = n_i + 1$.}, obtaining a distribution on $n_{i+1}$ qubits where adjacent states share the amplitude of the corresponding state in the $n_i$ qubit register. This effectively doubles the ``resolution'' of the $n_i$-qubit distribution without adding any additional information. We then obtain the correct $n_{i+1}$ qubit distribution by applying $\H$ a total of $t_{i+1}$ times\footnote{Similarly, $t_i$ refers to the number of iterations of $\H$ applied on the $i^{th}$ step of the procedure.}, we call these iterations ``correction'' iterations. We repeat this procedure until the desired distribution on the final number of qubits, $n_m$, is obtained. Thus, the procedure requires the calculation of a list $[t_1, t_2, ..., t_m]$ specifying the number of iterations of $\H$ to apply at each qubit count. As shown in SI~\ref{appendix:efficient_scaling_qubit_addition}, the final $n_m$ qubit variance obtained with any such list is given by,
\begin{align}
    \sigma^{2} = \frac{4^{n_m-1}-1}{12} + \frac{1}{4}\sum_{k=1}^{m} 4^{m-k}t_{k}.
\end{align}
Moreover, given a desired input variance on the real input grid $\hat{\sigma}^2$, it is straightforward to workout a schedule $[t_1, t_2, ..., t_m]$ producing the corresponding variance $\sigma^2$ in the computational basis.
In general, an effective initial setting for $t_i$ is $t_2=t_3=...=t_m=c$ for some small constant integer $c$ (for example, we often take $c=2$), and then setting $t_1$ to whatever value produces the final desired variance. Such an assignment produces the shortest circuits possible, as it maximizes the number of iterations performed on the smallest qubit count, where the fewest number of iterations of $\H$ has the greatest impact on the variance obtained. As such, we may compute $t_1$ with,
\begin{align}\label{eqn:t_1_computation}
t_1 \gets \frac{\sigma^2 - \frac{4^{n_m - 1}-1}{12}-\frac{c}{4}\sum_{k=2}^{m}4^{m -k}}{4^{n_m - 1}}.
\end{align}
However, when $t_1$ is not an integer some error is incurred as only an integer number of iterations can be applied,  and so some minor tuning of $t_i$ for greater values of $i$ may be required to get the exact desired variance (noting that it is always possible and straightforward to do so).
Finally, we shift the distribution (by adding a constant amount to all states) to obtain the correct mean. The shift may be implemented using a simple adder circuit, and the shift amount is calculated as the difference of the obtained mean,
\begin{align}
    \mu_{obtained} = \frac{1}{2}\left( 2^{n_m-1}-1 \right) +  \frac{1}{2}\sum_{k=1}^{m} 2^{m-k}t_k
\end{align}
(also derived in SI~\ref{appendix:efficient_scaling_qubit_addition}) and the desired input mean (in integer units).

The correctness of this procedure, as well as the analysis of the variance as a function of the number of correction iterations, is provided in SI~\ref{appendix:efficient_scaling_qubit_addition}. Furthermore, the error in this qubit-scaling approximation is entirely corrected, meaning that no additional error is introduced by this optimization. 

\subsubsection{Implementing \texorpdfstring{$\H$}{H}}
To ensure the efficiency of this procedure, it is essential that $\H$ be implementable with a polynomial-depth circuit. We have explored a number of approaches, which may be broken into two high-level categories: MCMR-based and MCMR-free. The MCMR-based approach requires only a single ancillary qubit to produce the desired distribution at the cost of requiring a constant-bounded number of expected circuit evaluations, as proven in SI~\ref{sec:aysmptotic_expected_trials}. The circuit for the MCMR-based approach is derived and presented in SI~\ref{section:h_mcmr_approach}. The MCMR-free approach requires a number of ancilla qubits scaling linearly in the number of applications of $\H$ required. This approach is also briefly discussed in SI~\ref{section:h_mcmr_approach}, and is discussed in more detail in SI~\ref{sec:mcmr_free_discussion}. The complexity of $\H$ is simply the complexity of the adder gate implementation used. A number of possible implementations for the adder gate are discussed in SI~\ref{appendix:adder_complexity}, but in summary, this work uses a QFT adder which requires $\mathcal{O}(n)$ depth (if already in Fourier space), and no additional ancilla qubits. Alternatively, to avoid the use of the QFT the adder may be implemented with $\mathcal{O}(n\log n)$ depth if $\mathcal{O}(n)$ ancilla qubits are available.

\subsubsection{Noise Resistance}
Finally, the primary variant of the algorithm, the MCMR-based approach, has provable resistance to hardware errors by virtue of the formulation of $\H$. This property is discussed in detail in SI~\ref{appendix:noise_resistance}, and follows from the probability of the ancilla qubit collapsing to the $\ket{1}$ state in each application of $\H$.
In summary, the $\ket{0}$ and $\ket{1}$ ancilla states are used to produce a non-unitary transformation in the main register. When the $\ket{0}$ ancilla state is measured, it is known that the operation was applied successfully. 
Of essence to the efficiency of this procedure is that the probability of measuring the $\ket{1}$ state decays exponentially in the number of iterations applied. The amplitudes produced on the $\ket{1}$ state cancel near-optimally, and so if any hardware errors occur, with limited exceptions, they reduce the amount of destructive interference, thus increasing the probability of measuring the ancilla in the $\ket{1}$ state, and thereby increasing the probability of discarding the error-affected execution.
Essentially, this follows from the fact that the cancellation property relies on the continuity of the distribution, and the occurrence of  errors (in any noise model) almost always results in the introduction of discontinuities that cause the destructive interference on the $\ket{1}$ state not to occur.
This discussion is treated more rigorously in the aforementioned entry in SI~\ref{appendix:noise_resistance}.

\subsection{The Procedure}

\begin{algorithm}[t]
\begin{algorithmic}
\Procedure{GenerateNormalDistribution}{$\hat{\mu}$, $\hat{\sigma}^2$, $x_0$, $l$, $n_1$, $n_m$}
    \State $N = 2^{n_m}$
    \State $\sigma^2 = \frac{N^2}{l^2} \hat{\sigma}^2$
    \State $\mu = \frac{N}{l}\hat{\mu}$
    \State $\ket{\psi}_{n_1} \gets \ket{0}_{n_1}$
    \State $t_2=t_3=...=t_m = c$
    \State $t_1 = \left(\sigma^2 - \frac{4^{n_m - 1}-1}{12}-\frac{c}{4}\sum_{k=2}^{m}4^{m -k} \right)/4^{n_m - 1}$ \Comment{I.e. Equation~\ref{eqn:t_1_computation}}
    \For{$i = 1$ to $m$}
        \For{$j=1$ to $t_i$}
            \State Create the new state $\ket{\phi}_{n_{i+1}}$ by applying $\left(I^{\otimes n_i}\otimes H\right) A^{+1}_{n_{i+1}}$ to $\ket{\psi}_{n_i}\otimes \ket{+}$.
            \State Measure qubit $n_{i+1}$ of state $\ket{\phi}_{n_{i+1}}$ in the standard basis
            \State Let $x \in \{0,1\}$ be the result of the measurement
            \State Let $\ket{\phi}_{n_{i}}$ be the resulting partial state
            \If{$x = 1$}
                \State Procedure failed, discard result, repeat procedure until success.
            \EndIf 
            \State $\ket{\psi}_{n_{i}} \gets \ket{\phi}_{n_{i}}$
        \EndFor
        \If{$i < m$}
            \State $\ket{\psi}_{n_{i+1}} \gets \ket{\phi}_{n_{i}}\ket{+}$
        \EndIf
    \EndFor
    \State $\alpha \gets \mu - \frac{1}{2}\left(2^{n_m - 1} - 1 + \sum_{k=1}^{n_m}2^{n_m - k}t_k \right)$
    \State $\ket{\psi}_{n_{m}} \gets A^{\alpha} \ket{\psi}_{n_{m}}$ \Comment{Shifts the produced mean to the target mean}
    \State\Return $\ket{\psi}_{n_{m}}$ \Comment{The procedure has succeeded}
\EndProcedure
\end{algorithmic}
\caption{Generating Normal Distributions with Mid-Circuit Partial Measurement.\\
\small{• We use the notation $\ket{\psi}_k$ to make explicit that $\ket{\psi}_k$ is a $k$-qubit state.}\\
\small{• Here $\ket{+}$ refers to a single qubit uniform superposition.}\\
\small{• Arrows are used to update or define quantum states.}
}
\label{algo:state_prep}
\end{algorithm}

We provide a pseudocode sketch of the procedure in Algorithm~\ref{algo:state_prep}; here we assume the use of mid-circuit partial measurements. Here, $n_1$ is the initial constant number of qubits on which the scaling procedure begins, while $n_m$ is the final target number of qubits upon which the distribution is to be prepared. The iteration counts $t_i$ can be defined much more flexibility than prescribed by this algorithm, but for simplicity we provide this recommended setting as it usually produces quite shallow circuits. Additionally, $c$ is some small constant number; $c=2$ often works well. The subscript on a ket denotes the number of qubits in that quantum state. Moreover, $A^{\alpha}$ is an adder gate that adds $\alpha$ to all states (as described in SI~\ref{appendix:adder_gate}), while $A^{+1}_{n+1}$ is a conditonal adder gate conditioned on qubit $n+1$ (as described in SI~\ref{appendix:pertinent_adder_gate}). As usual, $H$ represents a single-qubit Hadamard gate. Finally, as proven in SI~\ref{sec:aysmptotic_expected_trials} the algorithm's probability of success is constant bounded.

All together, we have shown how to produce arbitrary normal distributions in quantum registers, with asymptotic complexity logarithmic in the desired resolution of the distribution.
Our procedure requires a polynomial circuit depth and a constant number of expected circuit executions, and can efficiently load a normal distribution with arbitrary variance onto a quantum register -- all with provable resistance to hardware noise.

\section{Experimental Demonstrations}

\begin{figure*}[t]
\centering
{
    \includegraphics[width=0.85\linewidth]{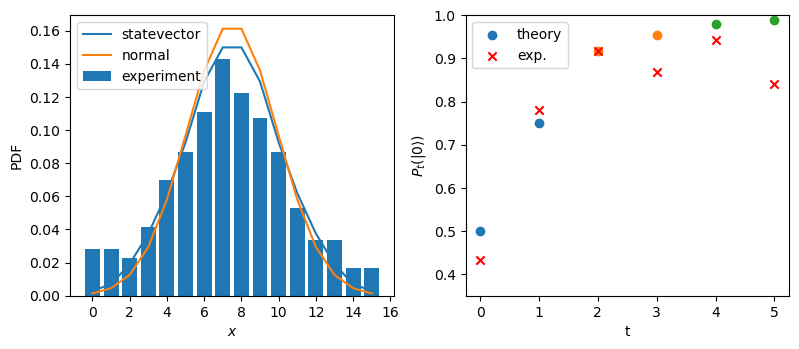}
}
    \caption{\textbf{Samples Drawn from a $5$-Qubit Implementation of the Algorithm on the Honeywell System Model H0.} In this experiment, the algorithm is configured with $n_1=2, n_2=3, n_3 =4$, and $t_1 = t_2 = t_3 = 2$. The distribution is then loaded into a quantum register using the proposed algorithm's MCMR variant, and then samples are drawn from the output. Thus, a total of $5$ qubits are used in the experiment. The resulting circuit has $75$ U1 gates, $60$ CX gates, $22$ U2 gates, and $10$ measurements. A total of 2500 samples are taken, of which only $532$ are kept in the algorithm's post-selection process. The left panel presents the post-selected results from the hardware-execution (bar-plot), the ideal statevector distribution (blue line), and the corresponding Gaussian distribution (orange line). Both plots are only assigned values directly on the integer grid points. The panel on the right compares the experimental and theoretical probabilities of obtaining the $\ket{0}$ state on the ancilla at the $t^{th}$ iteration, given that all preceding iterations yielded $\ket{0}$ ancillary measurements. The colors of the circles indicate different qubit scaling stages in the algorithm ($n_1$ - blue, $n_2$ - orange, $n_3$ - green).
    }
    \label{fig:exp2_demonstration}
\end{figure*}

\begin{figure*}[t]
\centering
{
    \includegraphics[width=0.825\linewidth]{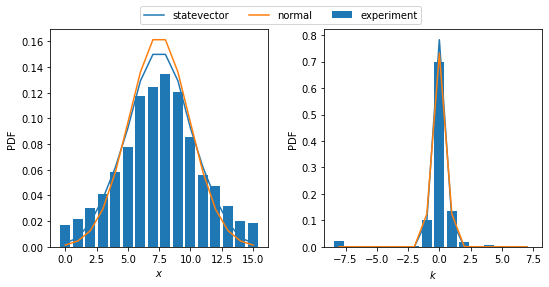}
}
    \caption{\textbf{Samples Drawn from a $5$-Qubit Implementation of the Algorithm on the Honeywell System Model H1-2 Measured in Two Bases, With Jump-Starting.} In both experiments, the algorithm is configured with $n_1=2, n_2=3, n_3 =4$, and $t_1 = t_2 = t_3 = 2$, and uses the MCMR variant (and thus requires one additional ancillary qubit). Additionally, in both experiments the two iterations at $n_1$ are performed with a deterministic procedure producing the exact two-qubit distribution, a process we call \textit{jump-starting}. 
    In the left panel, samples are drawn in the computational basis, while in the right panel samples are drawn in the Fourier basis.
    In the left panel, a total of $10000$ samples are taken, of which $7343$ are kept in post-selection. In the right panel, a total of $10000$ samples are taken, of which $6631$ are kept in post-selection.
    The blue-plot represents the ideal algorithm output in the absence of hardware noise, while the orange plot represent the exact corresponding normal distribution.
    }
    \label{fig:hardware_exp2}
\end{figure*}

In this section, we first present and analyze the results from running the MCMR version of the algorithm on the Honeywell System Model H0~\cite{Pino_2021} and Honeywell System Model H1-2, then we present the results comparing various variants of the algorithm in numerical simulations, and finally we present a discussion of the fidelity of the algorithm in theoretically ideal conditions.

\subsection{Quantum Hardware Experiments}

\begin{figure*}[t]
    \centering{
    \includegraphics[width=0.85\linewidth]{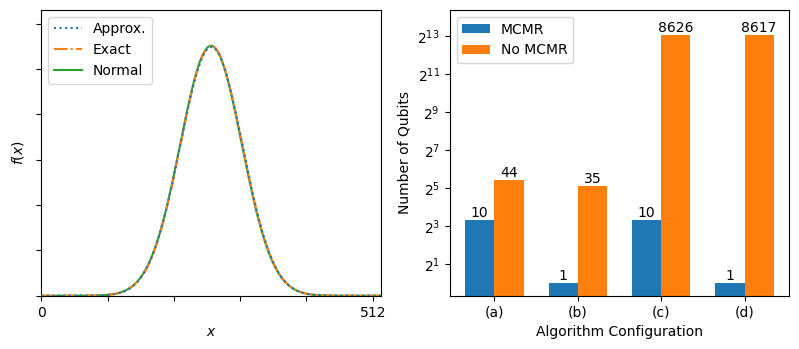}
    }
    \caption{\textbf{Simulated Experiment Comparing the MCMR-based and MCMR-free Approaches for the Exact and Approximate Methods.} The left panel of this figure displays the results obtained in theoretical simulation for the approximate method with $n_1=5$ scaling up to $n_5=9$, and with $t_1 = 32$ and $t_2=t_3=t_4=t_5=4$. The exact approach (i.e. no qubit scaling) is executed using $n=9$ qubits, and with an equivalent effective iteration count of $t=8617$. The normal distribution shown is produced by computing the variance and mean corresponding to $t=8617$, assuming that $l = N$. The panel on the right shows the total number of qubits and the number of ancillary qubits required to execute each configuration of the algorithm. The bars labelled $(a)$ represent the total qubit counts for the approximate algorithm, the bars labelled $(b)$ represent the ancillary qubit counts required for the approximate algorithm, the bars labelled $(c)$ represent the total qubit counts required for the exact algorithm, and the bars labelled $(d)$ represent the ancillary qubit counts required for the exact algorithm.}
    \label{fig:two_panel_demonstration}
\end{figure*}

The results from the hardware experiment are presented in Figure~\ref{fig:exp2_demonstration}, with the configuration described in the corresponding caption. We use an optimized version of the MCMR-based qubit-scaling algorithm, as presented in SI~\ref{section:main_quantum_circuit}. A total of $2500$ samples were taken in this experiment, among which $532$ occurred with all of the ancillary measurements yielding the $\ket{0}$ state. This corresponds to a post-selection rate of $21.28\%$.
In contrast, theoretical calculations indicate that performing this experiment in noiseless simulation would result in a post-selection rate of $\approx 31.93\%$. Therefore, there is approximately a $10\%$ deviation from the theoretical prediction and the experimental observations. In the right panel of Figure~\ref{fig:exp2_demonstration}, the y-axis shows the selection rate at the measurement of the $t^{th}$ application of $\H$. The selection rate at $t$ is defined as the probability of measuring a $\ket{0}$ at that specific iteration given that all prior measurements in the same experiment yielded $\ket{0}$.
In general, the selection rates observed on the ancillary qubit match the theoretical predictions, especially for the first three applications of $\H$, with the relatively small deviations well explained by the intrinsic stochastic variation given the limited quantity of samples collected.
After the third application of $\H$, the theoretical results and the experimental results start to differ more significantly, insofar as the experimental results reject more samples than error-free analysis predicts. With each application of $\H$, additional CX gates are executed, increasing the probability of an error occurring. As  predicted in SI~\ref{appendix:noise_resistance} as the probability of obtaining an error increases, the probability of discarding a circuit execution also increases, thus explaining the deviations observed. A similar argument also explains the $10\%$ experimental deviation from the theoretical prediction for the total number of shots kept. Indeed, this experimental deviation from the theoretical prediction supports the claim of the algorithm being noise resistant.

The left panel of Figure~\ref{fig:hardware_exp2} presents the results of the same experiment as Figure~\ref{fig:exp2_demonstration}, with the only difference being that the first two iterations of the procedure on $2$ qubits were jump-started (defined precisely shortly), and the circuit was executed on Honeywell System Model H1-2 instead of Model H0. We performed jump-starting to demonstrate the significant boost it can have on the percentage of circuits kept after post-selection, and we used Honeywell System Model H1-2 instead of Honeywell System Model H0 as it was the hardware available when running the experiment. The right panel of Figure~\ref{fig:hardware_exp2} presents the results of the same experiment as the left panel, only with an additional quantum Fourier transform applied at the end (which is actually implemented by removing the final inverse quantum Fourier transform, and appropriately reordering the bits). The purpose of the experiment shown in the right panel is to highlight the fact that the algorithm's resistance to hardware errors results in it not only drawing samples according to the correct probability distribution, but also actually produces the correct quantum state (with the correct relative phases, etc). 

We first discuss the left panel of Figure~\ref{fig:hardware_exp2}. A $k$-jump start removes the first $k$ applications of $\H$, and instead some other circuit (e.g. a universal circuit when $n_i$ is small) is used to exactly produce the state that would have otherwise been created. As the algorithm's probability of failure exponentially decreases after each application of $\H$, setting $k$ to a small value such as $2$ can significantly increase the probability of success, as it removes the iterations of the procedure where most failures would otherwise have occurred. Moreover, as the algorithm assumes that $n_1$ is some small constant number of qubits, such a jump-starting circuit has a small constant depth, and thus doesn't significantly increase the overall depth of the circuit. To ensure that the jump-starting circuit need only act on a small number of qubits, it is not recommended to use a value of $k$ greater than $t_1$.
The theory explaining the benefits of jump starting is clear, and in practice we obtain the post-selection rate of $73.42\%$ in the left panel of Figure~\ref{fig:hardware_exp2} (as opposed to the post-selection rate of $21.28\%$ in Figure~\ref{fig:exp2_demonstration}), neatly demonstrating the significant benefits the technique incurs. 
As an additional note, theoretical calculations indicate that in the absence of noise, a $2$-jump start in this experiment would yield a post-selection rate of $\approx 85.16\%$, approximately an $11.74\%$ deviation from the experimental results.

We now discuss the right panel of Figure~\ref{fig:hardware_exp2}. The purpose of this experiment is to demonstrate that our procedure not only produces samples from the correct probability distribution when executed on real quantum hardware, but also produces the correct amplitudes on the quantum states. In particular, this experiment exploits the fact that the Fourier transform of a normal distribution is another normal distribution, with a new mean and a new variance. If the normal distribution produced in the quantum register by the algorithm were correct and free of any phase errors (which would actually be bit-flip errors in Fourier space), then the resulting distribution after a final Fourier transform would be as shown by the blue and orange curves in the right panel of Figure~\ref{fig:hardware_exp2}. Indeed, sampling after this final Fourier transform yields a probability distribution matching that predicted by the theory, confirming that our algorithm is producing a state free of both bit-flip and phase-flip errors (and thus that we are not only producing the correct probability distribution, but also the correct amplitudes). 
Note that this is consistent with our proof that the algorithm is resistant to \textit{both} phase-flip and bit-flip errors.

\subsection{Simulated Experiment}

\begin{figure*}[t]
    \centering{
    \includegraphics[width=0.6\linewidth]{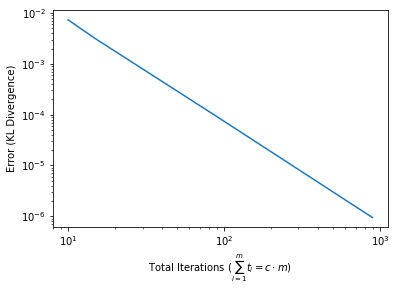}
    }
    \caption{\textbf{Theoretical Fidelity Analysis for Algorithm.} This figure shows the error between the distribution produced by the algorithm and the corresponding exact analytical normal distribution. Error is measured as the KL-Divergence between the produced distribution (treating the amplitudes as probabilities) and the exact distribution. The algorithm is configured with $n_1=6, n_2=7, n_3=8, n_4=9, n_5=10$, and with $t_1=t_2=t_3=t_4=t_5=c$ for $c$ an integer in the domain $[2, 180]$.
    }
    \label{fig:fidelity_analysis}
\end{figure*}

In Figure~\ref{fig:two_panel_demonstration}, we compare the exact and approximate techniques, along with their MCMR-based and MCMR-free variants. Here, we call the approach without qubit scaling the exact approach as it produces an exact Binomial distribution, and we call the qubit scaling approach approximate as it does not exactly produce a Binomial distribution (rather it just directly approximates a normal distribution through the central limit theorem).
We first configure the approximate algorithm with $n_1=5$ scaling up to $n_5=9$, and with $t_1 = 32$ and $t_2=t_3=t_4=t_5=4$. We run the exact algorithm on $n=9$ qubits, and with an iteration count of $t=8617$.  Both approaches are expected to generate a distribution with $\sigma^2 = 2154.25$ and $\mu = 256$. 
Results from the simulations are plotted along with a normal distribution with the same mean and variance.
All of the produced curves are shown on top of each other, demonstrating the effective equivalence of all three distributions (as expected, from a central limit theorem approximation when $N=2^9$). The panel on the right of Figure~\ref{fig:two_panel_demonstration} shows the number of ancillary and total qubits required for the exact and approximate simulators with and without MCMR. The exact simulator requires $8617\times$ fewer ancillary qubits and $862.6\times$ fewer total qubits when using MCMR. The approximate simulator requires $35\times$ fewer ancillary qubits, and $4.4\times$ fewer total qubits. Clearly, these reductions in required qubit resources will only increase as the variance of the desired distribution increases. 

\subsection{Fidelity Analysis}
In Figure~\ref{fig:fidelity_analysis} we conduct an analysis of the fidelity of the states produced by Algorithm~\ref{algo:state_prep} for a constant $c$ (in the absence of hardware error). Error is measured as the KL divergence between the distribution produced on the amplitudes and the corresponding exact normal distribution. Note that a similar plot is produced when error is measured as $1-\ip{\psi}{\phi}$ (where $\ket{\psi}$ is the exact distribution and $\ket{\phi}$ is the approximate distribution) only with all errors decreased by about one order of magnitude. 
We treat the amplitudes as probabilities for the purpose of this calculation, a valid assumption since the algorithm is restricted to the real plane, and each normal distribution in amplitudes is isomorphic to a normal distribution with a different variance in the corresponding probability space.
As stated in the caption, we configure the algorithm with $n_1=6, n_2=7, n_3=8, n_4=9, n_5=10$, and with $t_1=t_2=t_3=t_4=t_5=c$ for $c$ an integer in the domain $[2, 180]$. The plot shows a linear relationship between the log-error and log-iteration-count. Running a linear regression on the log-log data, we find a line of best fit with slope $-1.990334$, an r-value of $-0.999996$, and a p-value of $0.0$. This supports the conclusion that in the model where a constant $c$ iterations are performed at each qubit count, the error scales as $\Theta(\frac{1}{t^2})$ (where $t=\sum_{i=1}^m t_i = cm$). 
It is possible that this error dependence could be improved (potentially asymptotically) by allocating iterations at various qubit counts non-uniformly (i.e. not using a constant $c$).
However, it is not obvious how this would work, as the error from the central limit theorem approximation to the normal distribution depends on the number of random variables being summed (i.e. the number of applications of $\H$) and does not necessarily appear to depend on which qubit count an iteration of the procedure is applied. That is to say, the error of the procedure may simply depend on the total number of iterations performed, irrespective of the schedule for each $t_i$ that is selected. As such, to obtain a given variance, it may make sense to perform as many iterations as possible at the final qubit count $n_m$ so as to produce the desired fidelity (as iterations at the final qubit count have the least impact on the produced variance).
We leave such explorations as topics for future investigations. 

It is worth briefly explaining that the only source of error in the qubit-scaling procedure comes from the central limit theorem approximation to the normal distribution. To begin, as proven in the SI, we are clearly producing some distribution with the correct mean and variance (or as close as the selected discretization allows). Thus, the remaining source of error comes from how ``close'' the produced distribution is to being normal. Clearly, if it were exactly a normal distribution, there would be no error in the produced distribution (as it would be a normal distribution with the correct mean and variance). However, our distribution is only approximately normal (with the approximation coming from the central limit theorem as we are essentially summing a number of random variables) and thus the error is only that incurred by the central limit theorem approximation.

\subsection{Probability of Success}

When the algorithm is configured such that all of its $t$ iterations are performed at a single qubit count (i.e. without the qubit scaling procedure), the probability of success scales with $\Omega(t^{-1/2})$ as shown in SI~\ref{sec:aysmptotic_expected_trials_no_qs}. In the absence of the qubit scaling procedure, Equation~\ref{eqn:t_in_terms_of_var} clearly states the relationship between $t$ and the user-specified input parameters, allowing us to write this bound on the probability of success as $\Omega(\frac{l}{2^n \hat{\sigma}})$.

The analysis for the probability of success becomes more complicated when the qubit scaling procedure is utilized. The main insight required to understand the superior scaling of this approach comes from the observation that the probability of success of a \textit{single} application of $\H$ doubles when $\H$ is applied to the state $\ket{\psi}\ket{+}$ as opposed to just the state $\ket{\psi}$ (where the amplitudes of $\ket{\psi}$ are some discretization of a continuous distribution). As a result, each time the qubit count is increased by adding a qubit in the $\ket{+}$ state as the least-significant qubit, the probability of success of the next application of $\H$ doubles (meaning the probability of failure of any given iteration decays exponentially in the number of qubits added). Moreover, by assumption, $n_1$ is selected to be some small constant number of qubits, and thus any value of $t_1$ which we select will also be a small constant. 
Observing that we need only perform a constant-bounded number of iterations at each qubit count in the scaling procedure to obtain an arbitrary variance (since $t_1$ at $n_1$ gives the bulk of the ``shape'' of the distribution, and all $t_j$ with $j>1$ simply correct the error incurred from adding qubits in the $\ket{+}$ state) we have $t_2,...,t_m < c$ for some constant $c$ (and again $t_1$ is another constant). As a result, Equation~\ref{eqn:prob_success_qubit_scaling_final_bound} from  SI~\ref{sec:aysmptotic_expected_trials_qs} says that the expected number of trials before the procedure succeeds once, $E[T]$, is constant bounded as $E[T] < 2^{t_{\text{max}} + t_1}$ (where $t_{\text{max}} = \max_{2 \le  k\le m} t_k$) as $t_{\text{max}}$ and $t_1$ are both constant bounded.
Therefore, the cumulative probability of success is bounded from below as, $\Omega(\frac{1}{2^{c + t_1}})$. Since this is a constant, it is not necessary to provide this bound in terms of the user-specified input parameters.

To obtain an arbitrary normal distribution, we have shown that our approach requires a constant-bounded number of circuit executions (i.e. has a constant lower-bound on the success rate). However, we note that there is an apparent trade-off between the success-rate of the algorithm and the fidelity of the distributions obtained. 
In particular, as previously discussed, the fidelity of the algorithm roughly depends on the total number of times that $\H$ has been applied. However, our argument for the probability of success relies on the fact that we perform a constant-bounded number of iterations at each qubit count, limiting the ability of the user to fine-tune the fidelity of the obtained normal distribution. We simply observe that each additional iteration at the final qubit count $n_m$ has an exponentially small impact on the probability failure, and so the fidelity can be fine tuned by setting $t_m$ to a sufficiently large value (which we would not expect to substantially impact the overall probability of success) and then appropriately reducing $t_i$ for $i < m$.

Finally, we note one significant observation which can substantially improve the algorithm's overall probability of success. 
The probability of failure of an application of $\H$ decays exponentially in the number qubits added in the scaling procedure. As a result, the majority of the probability of failure comes from the first few iterations of the algorithm (namely from $t_1$). By assumption, $t_1$ is applied on a small constant number of qubits $n_1$, and as a result we can simply use a deterministic universal quantum state preparation procedure to prepare the exact $n_1$ qubit state that would have been produced by $t_1$ applications of $\H$~\cite{plesch2011quantum}. That is to say, we can use a universal quantum circuit to perform a $k$-jump start. Asymptotically, the cost of this procedure would roughly be $O(2^{n_1})$, which again is a constant.

\section{Conclusion}
This work presents a novel quantum algorithm for the generation of normal distributions in quantum registers, the latter of which runs with $\mathcal{O}(n^2)$ circuit depth (with $n$ being the number of qubits in the output quantum register) and a single ancilla, or with $\mathcal{O}(n \log n)$ circuit depth and $\mathcal{O}(n)$ ancilla qubits.
The algorithm uses a repeat-until-success scheme combined with MCMR technology, with a constant-bounded rate of success. Thus, to obtain a normal distribution with this approach, one need only run the procedure a constant number of times in expectation, independent of any of the user input parameters.
In addition, this work also demonstrates the potential of MCMR technology in the NISQ era by highlighting how it can enable existing qubit resources to be used asymptotically more efficiently in the introduction of non-unitary transformations, and by demonstrating how it allows qubit-efficient error detection and error mitigation techniques.

\section*{Acknowledgments}We would like to thank Saori Pastore (Washington University in St. Louis) for the insightful discussions on the material presented in this paper.  Special thanks to Tony Uttley, Brian Neyenhuis and the rest of the Honeywell Quantum Solutions team for assisting us on the execution of the experiments on the Honeywell System Model H0.

\section*{Disclaimer}This paper was prepared for information purposes by the Future Lab for Applied Research and Engineering (FLARE) group of JPMorgan Chase Bank, N.A..  This paper is not a product of the Research Department of JPMorgan Chase \& Co. or its affiliates.  Neither JPMorgan Chase \& Co. nor any of its affiliates make any explicit or implied representation or warranty and none of them accept any liability in connection with this paper, including, but limited to, the completeness, accuracy, reliability of information contained herein and the potential legal, compliance, tax or accounting effects thereof. This document is not intended as investment research or investment advice, or a recommendation, offer or solicitation for the purchase or sale of any security, financial instrument, financial product or service, or to be used in any way for evaluating the merits of participating in any transaction.

\section* {Author Contributions}
A.G.R. conceived of the project and designed the proposed algorithms. Y.S., A.G.R., P.M., and M.P. contributed further optimizations to the algorithms, and developed the theory. A.G.R., Y.S., and M.P. implemented the algorithm. All authors contributed to the manuscript. 

\printbibliography

\appendix

\section*{Supplementary Information: \\\vspace{0.3cm} Approximate and Exact Quantum Simulation of Galton Machines}

\section{Motivating the Algorithm}\label{appendix:algo_intuition}

\begin{figure}[t]
\centering
    {
    \includegraphics[width=0.25\linewidth]{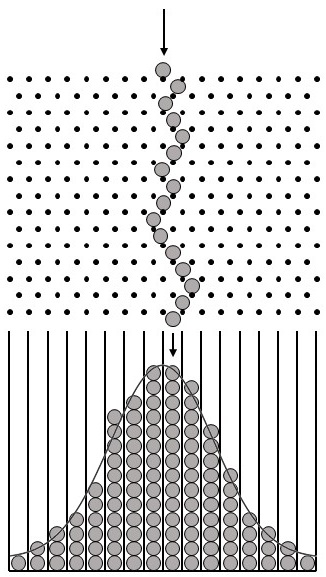}}
    \caption{\textbf{Galton Machine Visualization.} Visual demonstration of a Galton machine generating a normal distribution. Figure adapted from \textit{Option Pricing and Volatility}, by Sheldon Natenberg~\cite{Natenberg2014}.}
    \label{fig:galton_machine}
\end{figure}

Continuing with the notation introduced in Section~\ref{sec:algorithm}, we now provide the intuition motivating the state generating algorithm. According to the de Moivre–Laplace theorem, a binomial distribution $B(t, p)$ converges to a normal distribution with mean $tp$ and standard deviation $\sqrt{tp(1 - p)}$ as $t\to \infty$. This property is exploited by Galton machines, such as the one shown in Figure~\ref{fig:galton_machine}, to generate normal distributions by dropping balls through a sequence of rows, where in each row the falling ball may have its position shifted by either one bin to the left, or one bin to the right. As a result, the transition dynamics of a single row in a Galton machine may be described by the transition matrix $\H$ defined as (up to a normalization factor)
\begin{align*}
    \H\ket{j} = \ket{j - 1} + \ket{j + 1}.
\end{align*}
As previously mentioned, the quality of the approximation of the normal distribution produced by a Galton machine improves as more balls are dropped through the system. Moreover, as our quantum Galton approach simulates an infinite number of balls falling through the system (by virtue of operating on superpositions of ``bins''),  we avoid this type of error entirely. As a result, combined with the exponential growth in the number of bins in terms of the number of qubits, the distributions we produce are indistinguishable from the target normal distributions (i.e. our only source of error is that incurred by the central limit theorem approximation, and the central limit theorem error vanishes exponentially in the number of qubits, and so our approximation rapidly becomes exact). 

It is worth briefly mentioning that we explored another variant of the algorithm simulating $e^{\H t}$ (instead of $\H^t$) to obtain the desired normal distribution. This approach utilizes Imaginary Time Evolution (ITE) to simulate $e^{\H t}$, as proposed by McArdle \textit{et al.} in 2019~\cite{mcardle2019variational}, and is discussed in greater depth in SI~\ref{appendix:imaginary_time_evolution}.

\section{Implementing \texorpdfstring{$\H$}{H} as a Quantum Circuit with Mid-Circuit Measurement and Reuse (MCMR)}\label{section:h_mcmr_approach}
In order to make the implementation of $\H$ as a quantum circuit more straight forward, we will redefine the transition matrix. Instead of performing the transformation $\H\ket{j} =\ket{j-1} + \ket{j+1}$, we will now perform the mapping $\H\ket{j} = \ket{j} + \ket{j + 1}$. In so doing, we must pay additional attention to ensure that the mean of the generated distribution is configured correctly, and in exchange, we reduce the complexity of certain sections of the analysis. First, observe that this transition matrix still generates a normal distribution in the same way as the first matrix, only that instead of alternating between zero amplitude even and odd states with each application of $\H$, all states are utilized without the need for the input state to contain a superposition of adjacent states (or without the need of additional ancillary qubits). We will now define an addition operator, $A^{+1}_{n+1}$, where the subscript indicates the control qubit, and the superscript indicates the quantity added to each state in the first $n$-qubit register. For example, $A_{n+1}^{+1}\ket{j}\ket{1} = \ket{j + 1}\ket{1}$, and $A_{n+1}^{+1}\ket{j}\ket{0} = \ket{j}\ket{0}$. We then derive the quantum circuit as follows.
First we apply a Hadamard gate, $H$, on the ancilla qubit,
\begin{align*}
    I^{\otimes n}\otimes H\ket{j}\ket{0} =& \ket{j}(\ket{0} + \ket{1}).
\end{align*}
Then we apply the $+1$ gate on the main register, controlled on the ancilla,
\begin{align*}
    A^{+1}_{n + 1} \ket{j}(\ket{0} + \ket{1}) =& \ket{j, 0} + \ket{j + 1, 1}.
\end{align*}
Finally, we apply another Hadamard gate on the ancilla,
\begin{align*}
    &I^{\otimes n}\otimes H(\ket{j, 0} + \ket{j + 1, 1}) \\
    &= \ket{j, 0} + \ket{j, 1} + \ket{j + 1, 0} - \ket{j + 1, 1}\\
    &= (\ket{j} + \ket{j + 1})\ket{0} + (\ket{j} - \ket{j + 1})\ket{1}.
\end{align*}
We see that upon measuring the ancilla qubit in the $\ket{0}$ state, the main register will contain the state $\ket{j} + \ket{j + 1}$ as desired. However, if we measure the ancilla in the $\ket{1}$ state, the main register will be found in the incorrect state $\ket{j} - \ket{j + 1}$. Initially, methods for correcting this error were explored, however, we realized that $\ket{j} - \ket{j + 1}$ actually represents a pattern of destructive interference that exponentially approaches net zero amplitude as a function of the number of iterations applied. As a result, a single iteration of the circuit implementation of $\H$ may be produced by following the procedure just described, and by measuring the ancilla qubit, continuing if the desirable ancilla state is measured, and terminating execution if the undesirable state is measured. As will be proved shortly, this procedure requires an increase in the expected number of circuit evaluations growing sub-linearly in the number of iterations performed, and therefore results in a polynomially bounded total number of circuit executions for the algorithm as a whole. These claims are motivated in SI~\ref{subsubsec:cancellation_analysis}, and proven in SI~\ref{sec:aysmptotic_expected_trials}. 
In contrast, in the MCMR-free approach, as the single ancilla is not reused subsequently to being measured, a new ancilla qubit must be added for each application of $\H$. Each ancilla may be measured at the end of the application of $\H$, or by the principle of deferred measurement~\cite{nielsen2002quantum}, all added ancillas may be simultaneously measured at the end of the circuit's execution. A more detailed explanation of the MCMR-free approach is presented in SI~\ref{sec:mcmr_free_discussion}.

\subsection{Ancilla \texorpdfstring{$\ket{1}$}{|1>} State Cancellation Analysis}\label{subsubsec:cancellation_analysis}

\begin{figure}
\centering
    {
    \includegraphics[width=0.9\linewidth]{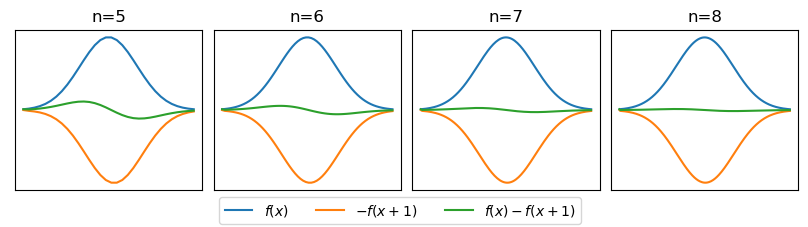}}
    \caption{\textbf{Amplitude Cancellation Demonstration.} This figure demonstrates the interference pattern produced when an initial distribution $f(x)$ is mapped to $f(x) - f(x + 1)$ at various qubit counts.}
    \label{fig:measure_ancilla_interference}
\end{figure}

The implementation of $\H$ just described relies upon performing a partial measurement on the state, $(\ket{j} + \ket{j + 1})\ket{0} + (\ket{j} - \ket{j + 1})\ket{1}$. If the ancilla is measured in the $\ket{0}$ state, the desired transformation yielding $\ket{j} + \ket{j + 1}$ has been obtained in the main register. If the ancilla is measured in the $\ket{1}$ state, the undesired state of $\ket{j} - \ket{j + 1}$ is obtained. We claim, and prove in SI~\ref{sec:aysmptotic_expected_trials}, that the probability of measuring the $\ket{1}$ ancillary state rapidly vanishes as a function of the number of applications of $\H$ performed, and that as a result, we only need to repeat the algorithm's execution a number of times sub-linear in the number of applications of $\H$ (and in the case of the qubit scaling approach, a constant-bounded number of times). To understand this, it may be more clear to understand an equivalent statement. Given a distribution on $n$ qubits obtained by computing $\H^t$, with $t$ computed taking into account $n$ and the input variance, the behavior of the mapping $\ket{j} \to \ket{j} - \ket{j + 1}$ may be similarly expressed in terms of functions of the amplitudes as $f(x_j) \to f(x_j) - f(x_j + \Delta x)$, where $\Delta x = \frac{l}{2^n}$. As such, given a fixed variance $\hat{\sigma}^2$, as we increase the number of qubits, we are exponentially decreasing the relative shift between distributions $f(x_j)$ and $-f(x_j + \Delta x)$, and so the two distributions rapidly approach $f(x_j)$ and $-f(x_j)$, meaning that the interference pattern on the $\ket{1}$ ancilla state exponentially approaches complete destructive interference as the number of qubits increase. Indeed, in SI~\ref{appendix:general_1_state_cancellation} we have shown that this cancellation property holds for any continuous function $f$. This principle is clearly illustrated in Figure~\ref{fig:measure_ancilla_interference}, where the interference pattern produced rapidly approaches complete destructive interference as the number of qubits increase up to $8$. 

\subsection{Quantum Circuit Implementation of \texorpdfstring{$\H$}{H}} 
\begin{figure}
\centering
    {
    \includegraphics[width=0.5\linewidth]{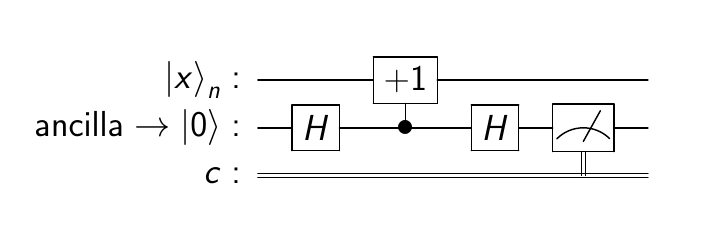}}
    \caption{Quantum circuit implementation of $\H$, with the definition $\H\ket{j} = \ket{j} + \ket{j + 1}$ (up to a normalization factor). Note that circuit execution only proceeds if the ancilla is measured in the $\ket{1}$ state.}
    \label{fig:H_one_iter}
\end{figure}

Again, the mapping we wish to perform is given by $\H\ket{j} = \ket{j} + \ket{j + 1}$. To perform such a mapping, we use four components: the QFT, the (Fourier space) adder gate, the Hadamard gate, and mid circuit measurement and reuse. Assuming we are in the Fourier space, (i.e. a QFT has already been performed previously in the circuit), $\H$ is implemented as shown in Figure~\ref{fig:H_one_iter}. As the adder gate may be implemented with $O(n)$ two-qubit gates and depth, one iteration according to $\H$ is also implemented in linear depth and with linear two-qubit gates. The adder gate and its quantum circuit, along with the corresponding controlled adder gate, is presented in SI~\ref{appendix:pertinent_adder_gate} and SI~\ref{appendix:adder_gate}.

\subsection{Algorithm Quantum Circuit Implementation}\label{section:main_quantum_circuit}
\begin{figure*}[t]
\centering
    {
    \includegraphics[width=1\linewidth]{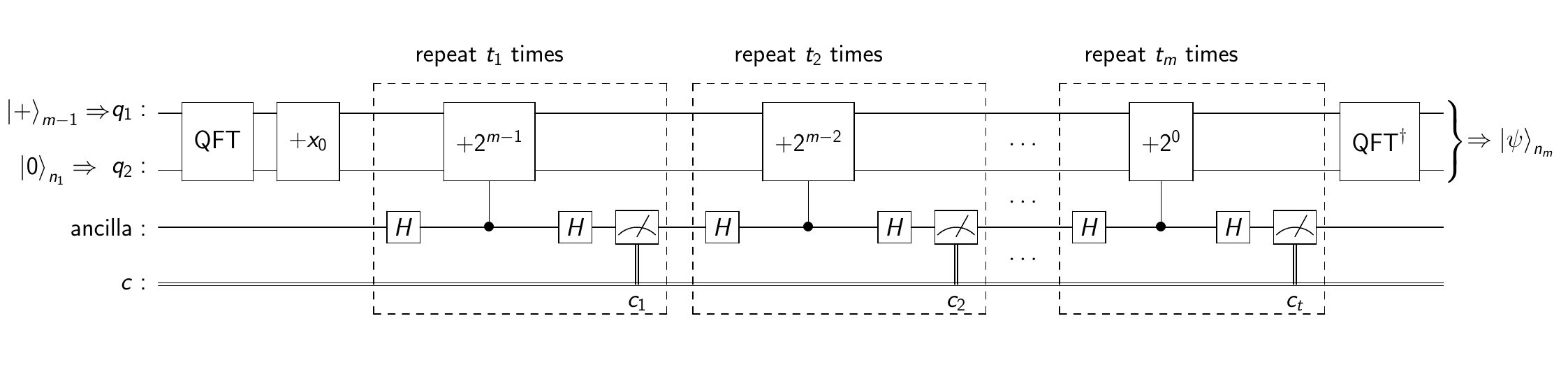}}
    \caption{Quantum circuit implementing the quantum Galton machine with qubit scaling. }
    \label{fig:quantum_circuit_scaling}
\end{figure*}

We first present a basic circuit used to implement $\H^t$, proceed to give the implementation of the full quantum circuit, including the qubit scaling procedure. 
The naive circuit (which is used for the exact simulation of Galton machines) uses two quantum registers: one with $n = n_m$ qubits for the output and the other $1$-qubit register as the ancilla. The two registers are initialized in joint state $\ket{0}_{n+1}$.
We then apply QFT on the $n$-qubit register to bring the quantum register into Fourier space.
Subsequently, we apply $\H$ which consists of a Hadamard gate on the ancilla qubit, followed by a controlled $+1$ gate ($A_{j}^{+1}$) controlled on the ancilla, a second Hadamard gate on the ancilla, and a measurement on the ancilla. 
The application of $\H$ is repeated $t$ times, with the value of $t$ given by Equation~\ref{eqn:t_in_terms_of_var}.
In the end, we apply an inverse-QFT on the output register to go back to the real space.
The success of the circuit depends on the measurement results on the ancilla qubit.
We repeat the circuit until all measurement results come out as $0$.

As discussed at the beginning of Section~\ref{sec:algorithm}, the naive implementation without qubit scaling described above would require exponential number of $\H$ applications, which renders the algorithm inefficient.
To address this issue, we use a qubit scaling scheme which significantly reduces the circuit depth with only nominal sacrifices in accuracy.
The circuit is illustrated in Figure~\ref{fig:quantum_circuit_scaling}.
We start with a small, constant, number of qubits $n_1$, and apply $\H$ $t_1$ times to get close to the desired variance of the distribution.
Then we add one qubit to the least significant end in the sate $\ket{+}$, and apply $\H$ $t_2$ times.
We repeat the process of adding qubits and applying $\H$ untill we reach the desired number of qubits $n_m$.
The values of $t_1, t_2, \dots, t_m$ are determined by the scaled variance $\hat{\sigma}^2$ of the distribution, the formulas of which are given in SI~\ref{appendix:efficient_scaling_qubit_addition}.
Note that the total number of applications of $\H$, i.e. $\sum_{r=1}^m t_r$ scales as $O(n_m)$, as discussed at the beginning of Section~\ref{sec:algorithm}, and in SI~\ref{appendix:efficient_scaling_qubit_addition}.

To avoid going back and forth between the real space and the Fourier space when increasing the number of qubits, we perform QFT on the entire $n_m$ qubit register and inverse-QFT at the very end.
Accordingly, the controlled $+1$ gates in the $r$-th stage with $n_r$ qubits would be changed to a controlled $+2^{m-r}$ gate on the $n_m$ qubit register.
Note that this still amounts to $n_r$ controlled-$U1$ operations in the controlled adder gate, as $U1$ gates with angles that are multiples of $2\pi$ amount to the identity gate.

In the case where the mean of the outcome distribution needs to be adjusted, an additional adder gate $+x_0$ may be added on the $n_m$-qubit register in the Fourier space, as shown in Figure~\ref{fig:quantum_circuit_scaling}.

\section{Implementing \texorpdfstring{$\H$}{H} as a Quantum Circuit without Mid-Circuit Measurement and Reuse (MCMR-free)}\label{sec:mcmr_free_discussion}
We now describe the implementation of the transition matrix $\H$ defined as $\H\ket{j} = \ket{j} + \ket{j + 1}$ without the use of Mid-Circuit Measurement and Reuse technology. We first perform a derivation similar to that performed in SI~\ref{section:h_mcmr_approach}. Examining one application of $\H$,
\begin{equation*}
    \ket{j}\ket{+} \xrightarrow{{A_{n+1}^{+1}}} \ket{j}\ket{0} + \ket{j + 1}\ket{1} \xrightarrow{I^{\otimes n}\otimes H} (\ket{j} + \ket{j + 1})\ket{0} + (\ket{j} - \ket{j + 1})\ket{1}.
\end{equation*}
When mid-circuit measurement is used, the preceding final quantum state may simply be measured, as described in a preceding SI entry. However, when MCMR technology is not available, the ancilla cannot be measured until the end of the circuit execution, and so to perform another iteration by $\H$ we must add another ancillary qubit so as to preserve the state of the first ancilla. In order to reveal some of the properties of this MCMR-free approach, we will now examine the state of the system after performing a second application of $\H$. First, we add another ancillary qubit initialized in the $\ket{+}$ state to obtain,
\begin{align*}
    (\ket{j} + \ket{j + 1})\ket{0}\ket{+} + (\ket{j} - \ket{j + 1})\ket{1}\ket{+}.
\end{align*}
We subsequently apply the adder $+1$ gate conditioned on the second ancilla, $A^{+1}_{n+2}$, to obtain the state,
\begin{align*}
    (\ket{j} + \ket{j + 1})\ket{00} + (\ket{j + 1} + \ket{j + 2})\ket{01} + (\ket{j} - \ket{j + 1})\ket{10} + (\ket{j + 1} - \ket{j + 2})\ket{11}.
\end{align*}
Simplifying and applying $I^{\otimes n + 1}\otimes H$ then yields,
\begin{align*}
    &(\ket{j} + 2\ket{j+1} + \ket{j + 2})\ket{00} + (\ket{j} - \ket{j + 2})\ket{01}\\ &+ (\ket{j} - \ket{j + 2})\ket{10} - (\ket{j} - 2\ket{j + 1} + \ket{j + 2})\ket{11}.
\end{align*}
It now becomes clear that upon adding a new ancillary qubit, and performing the standard operations required to implement an additional iteration according to $\H$, that the probability of measuring the $\ket{1}$ state of any preceding ancilla qubits remains unchanged. As such, the analysis regarding the probability of measuring all ancilla qubits in the $\ket{0}$ state, and thus of obtaining the correct state in the primary register, follows the same analysis as the MCMR-based version of the algorithm.

\subsection{Adder Gate Complexity}\label{appendix:adder_complexity}
All approaches require the implementation of an adder-plus-one gate ($A^{+1}$), which can be implemented either with additional ancillary qubits, or with no additional ancillary qubits. In the NISQ context, Table \ref{tab:QFT_depth} and Table \ref{tab:adder_depth} will provide estimate on the constant factor of the asymptotic circuit depth for different approaches described for the QFT and adder gate respectively. Note as we already know one of the two number in the additions, further optimization could be found in the described methods.
When not using ancillary qubits, the adder-plus-one gate may be implemented in Fourier space (requiring a  Quantum Fourier Transform (QFT) at the start of the overall circuit, and an inverse-QFT at the end) resulting in $\mathcal{O}(n)$ CX gates per application of $\H$~\cite{draper2000addition}. Alternatively, without additional ancillas, the adder-plus-one gate may be implemented with $\mathcal{O}(n^2)$ CX gates per application of $\H$. As a result, we use the QFT adder approach throughout this document, resulting in a circuit depth of $\mathcal{O}(n^2 + nt)$, which due to the qubit-scaling procedure already discussed is bounded by $\mathcal{O}(n^2)$. Alternatively, Draper \textit{et al.} present an adder requiring $\mathcal{O}(n)$ ancilla qubits, and scaling with $\mathcal{O}(\log n)$ circuit depth, without the need to enter the Fourier space, enabling our algorithm to scale with $\mathcal{O}(t \log n)$ circuit depth, if $\mathcal{O}(n)$ ancilla qubits are available~\cite{draper2004logarithmic}. 
One could adapt this space-depth trade-off to the specific hardware used. This reduction in depth would not only be seen on total complexity but also asymptotic complexity if we were to consider the approximate QFT \cite{nam2020approximate}. In a fault-tolerance context and with an error $\epsilon$, we would require for the adder and QFT, $\mathcal{O}((nt + n^2)\log(1/\epsilon))$ gates or just $\mathcal{O}((nt + n)\log(1/\epsilon))$ gates using the approximate QFT \cite{goto2014resource}. 

\begin{table}[h]
    \centering
    \begin{tabular}{c||c|c|c|c|c|c|}
         Component &  QFT &  Approximate QFT\\
         \hline
         Depth & $\mathcal{O}(n^2)$ &$\mathcal{O}(n\log n)$   \\
         Constant estimate & $2$ & $30$ \\
         Number of qubits & $n$ & $n+3\left \lceil{\log n}\right \rceil -4$ \\
         
    \end{tabular}
    \caption{QFT in NISQ context}
    \label{tab:QFT_depth}
\end{table}
\begin{table}[h]
\centering
    \begin{tabular}{c||c|c|c|}
         Component & Phase Adder & Draper\\
         \hline
         Depth & $\mathcal{O}(t n)$ & $\mathcal{O}(t\log n)$ \\
         Constant estimate & $2$ & $4$ \\
         Number of qubits & $2n$ & $3n$ \\
         
    \end{tabular}
    \caption{$t$ adders in NISQ context}
    \label{tab:adder_depth}
\end{table}

\subsection{The Adder Gate}\label{appendix:pertinent_adder_gate}

\begin{figure}[b]
\centering
    {
    \includegraphics[width=0.7\linewidth]{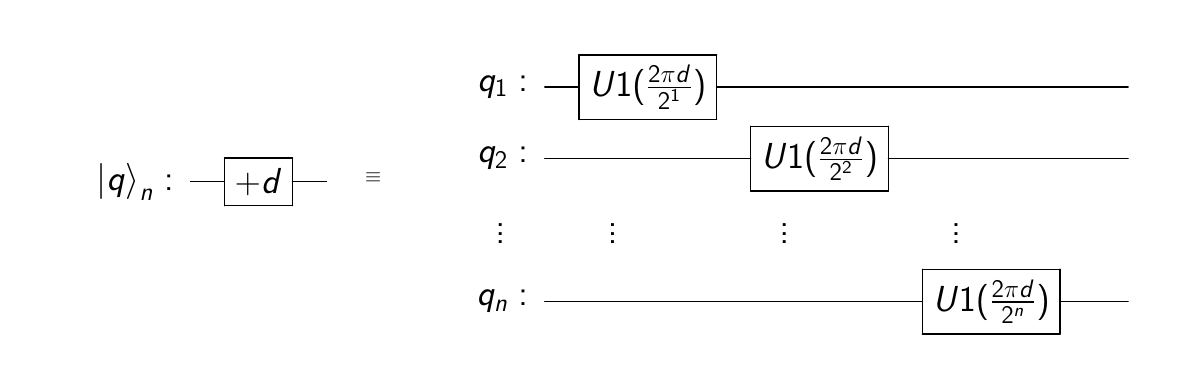}}
    \caption{Quantum circuit implementing the Fourier space adder gate $+d$. }
    \label{fig:qft_adder_circuit}
\end{figure}

\begin{figure}[t]
\centering
    {
    \includegraphics[width=0.7\linewidth]{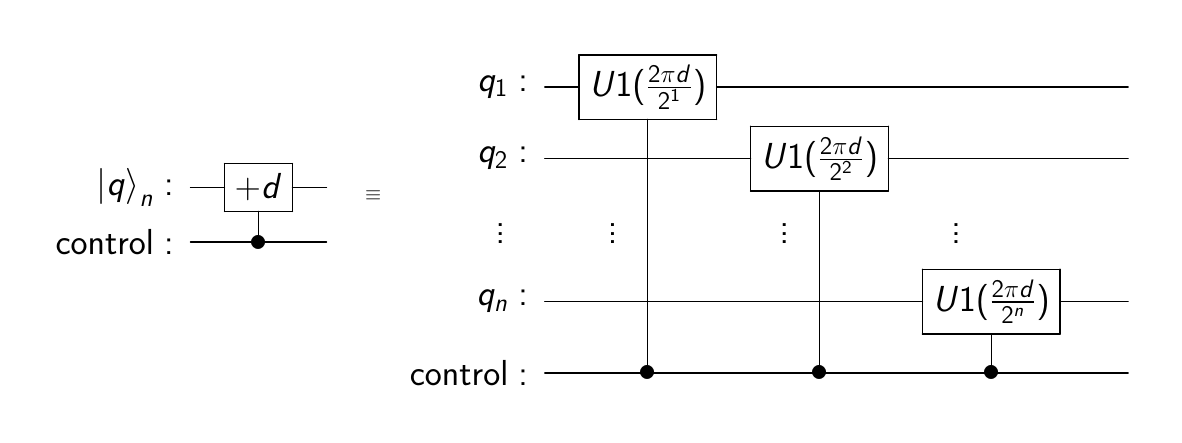}}
    \caption{Quantum circuit implementing the Fourier space controlled adder gate $A_{\text{control}}^{+d}$. }
    \label{fig:qft_ctrl_adder_circuit}
\end{figure}

The circuit implementation of the $+d$ gate is shown in Figure~\ref{fig:qft_adder_circuit}.
The general form of the $+d$ gate, with $d$ being an integer, is $\bigotimes_{k=1}^{n} U1(2\pi {d}/{2^k})$, where $n$ is the number of qubits in the main register~\cite{beauregard2002circuit}.
Moreover, as our adder gates are conditioned on the state of the ancillary qubit, the general form of the controlled adder gate $A_{j}^{+d}$ is given by,
\begin{align*}
    A_{j}^{+d} = \prod_{k=1}^{n} U1_j^{k}(2\pi \frac{d}{2^k} ),
\end{align*}
where a $U1_{a}^b$ gate is a $U1$ gate with a control on qubit $a$, a target on qubit $b$, and identity on all other qubits. 
Moreover, a U1 gate is defined as follows,
\begin{align*}
    U1(\lambda) =
    \begin{pmatrix}
        1 & 0\\
        0 & e^{i \lambda}
    \end{pmatrix},
\end{align*}
and is equivalent to an $R_z(\lambda)$ gate up to a global phase of $e^{i\frac{\lambda}{2}}$, as defined by Barenco \textit{et al.} in 1995~\cite{barenco1995elementary}.
The circuit for $A_{j}^{+d}$ is shown in Figure~\ref{fig:qft_ctrl_adder_circuit}.
For a more comprehensive discussion of the adder gate, see SI~\ref{appendix:adder_gate}.

\subsection{The General Adder Gate}\label{appendix:adder_gate}
Let $x \in \{0, 1\}^n$. When applying the QFT to such a binary-labelled standard basis vector, we obtain:
\begin{align*}
    \frac{1}{2^{n/2}}\bigotimes_{l=0}^{n-1}( \ket{0} + e^{2\pi ix\frac{2^l}{2^n}}\ket{1}).
\end{align*}
Thus, the action of the +1 gate is given by:
\begin{align*}
    &\frac{1}{2^{n/2}}\bigotimes_{l=1}^{n}U1(2\pi i \frac{2^l}{2^n})( \ket{0} + e^{2\pi ix\frac{2^l}{2^n}})\ket{1}\\
    &=\frac{1}{2^{n/2}}\bigotimes_{l=1}^{n}\left[ \ket{0} + \exp\left({2\pi i(x+1)\frac{2^{l}}{2^n}}\right)\ket{1}\right].
\end{align*}
A subsequent application of the inverse QFT would yield the desired state of $\ket{x+1}$. Note that multiple adder gates may be chained together while in the amplitude space, meaning that only one QFT would be required at the start of the circuit, and only one inverse QFT is requried at the end of the circuit. It is also worth noting that the periodicity of the imaginary exponential implies that this addition is actually addition modulo $2^n$, and so our $\H$ operation encounters a boundary condition when there are non-zero amplitude states $\ket{x=0}$ or $\ket{x = 2^n}$ (in practice, the boundary condition only becomes significant when the amplitudes of these extreme states are much greater than zero). However, it would not be expected for any algorithm to be able to circumvent this limitation, as it is intrinsic to the finite quantum mechanical system. Finally we can model a $A^{+k}$ gate (adder $+k$) as follows, 
\begin{align*}
    A^{+k} = \bigotimes_{l=1}^{n}U1(2\pi i \frac{2^l}{2^n} k).
\end{align*}

\section{Efficient Scaling Through Qubit Addition}\label{appendix:efficient_scaling_qubit_addition}
\begin{figure}
    {\centering
    \includegraphics[width=1\linewidth]{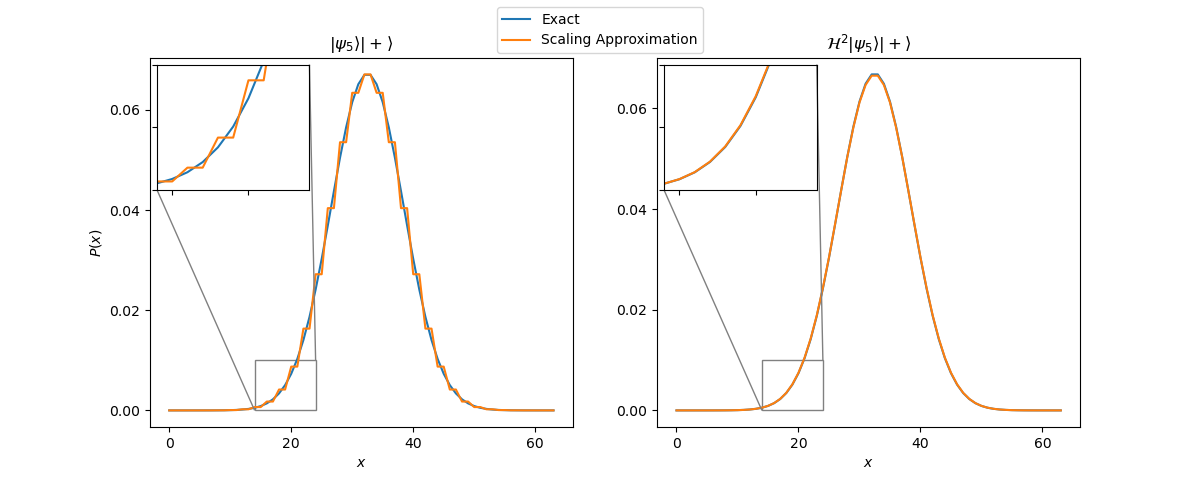}}
    \caption{\textbf{Demonstration of Correction Iterations in the Efficient Scaling Procedure.} Two Gaussians generated by the algorithm on $6$ qubits are shown. The scaling approximation curve in the left panel is obtained with $t_1=70$ and $t_2=0$, meaning that no correction iterations are performed after the scaling qubit is added to the least significant bit. The scaling approximation curve in the right panel is obtained with $t_1=70$ and $t_2=2$, meaning that two correction iterations are applied after the scaling qubit is added. In both panels, the exact curve is generated by applying $t_\text{eff}$ iterations on \textit{only} the final qubit count (meaning that the scaling procedure is not used), where $t_\text{eff} = 281$ in the panel on the left, and $t_\text{eff}=283$ in the panel on the right.}
    \label{fig:qubit_scaling_step_visualization}
\end{figure}

As discussed, if we wish to execute $\H^t$ on the quantum processor, naively we would expect that to obtain a given variance we require a number of applications of $\H$ that grow exponentially in the number of qubits in the final quantum register, $n_{m}$, in accordance with Equation~\ref{eqn:t_in_terms_of_var}. However, instead of performing all $t$ iterations in the space of $n_{m}$ qubits, we can produce a distribution with the same variance $\hat{\sigma}^2$ on a small, constant number of qubits. In doing so, we essentially produce a ``low-resolution'' version of the distribution that has the overall correct shape of the final distribution. We subsequently add a qubit in the $\ket{+}$ state to the least significant qubit, apply a number of correction iterations implemented directly as the $\H$ operator, and repeat this addition and iteration procedure $n_{m} - n_{1}$ times. To understand how applying $\H$ some number of times increases the resolution of the distribution after adding the $\ket{+}$ qubit, it is important to better understand the distribution obtained after the addition. Figure~\ref{fig:qubit_scaling_step_visualization} demonstrates both the distribution obtained immediately after adding a qubit in the $\ket{+}$ state (in the panel on the left), and the distribution obtained after adding a qubit in the $\ket{+}$ state followed by performing two correction iterations (directly implemented as applications of $\H$). As expected, by virtue of mapping the amplitude of state $\ket{x}$ to states $\ket{2x}$ and $\ket{2x + 1}$, each pair of two states shown in the figure have the same amplitude, resulting in the step-pattern shown. The panel on the right illustrates that it only takes two correction iterations to almost perfectly match the exact distribution. This is consistent with the theory discussed shortly. Moreover, while not shown in the figure, applying a couple more correction iterations would make the scaling approximation and exact distributions completely indistinguishable. 

First, we define the $i$-qubit grid as, $\{X_i\}$. Then, the grid on $i+1$-qubits (obtained simply by adding a qubit in the $|+\rangle$ state) is given by, $\{2X_i, 2X_i + 1\}$. Allow the $i$-qubit mean to given by $\mu_i$. Upon doubling the number of qubits (with adjacent even and odd states sharing their amplitudes), the mean becomes $\mu_{i+1} = 2\mu_i + 0.5$. From the definition of variance, we have
\begin{align*}
    \sigma_i^2 = \sum_{x\in X_i}(x - \mu_i)^2 P(x).
\end{align*}
Thus we can compute the variance of the resulting distribution with,
\begin{align*}
    \sigma^2_{i + 1} &= \sum_{x \in X_{i + 1}}(x - \mu_{i + 1})^2P(x)\\
    &= \frac{1}{2}\left(\sum_{x \in X_{i}}(2x - \mu_{i + 1})^2P(x) + \sum_{x \in X_{i}}(2x + 1 - \mu_{i + 1})^2P(x)\right)\\
    &= \frac{1}{2}\sum_{x \in X_{i}}\left[(2x - 2\mu_i - 0.5)^2 + (2x - 2\mu_i + 0.5)^2 \right]P(x) \\
    &= \sum_{x \in X_{i}}\left[4(x - \mu_i)^2 + 0.5^2 \right]P(x) \\
    &= 4\left[\sum_{x \in X_{i}}(x - \mu_i)^2 P(x)\right] + 0.5^2,
\end{align*}
Therefore,
\begin{align}\label{eqn:variance_qubit_scaling_relation}
    \sigma^2_{i+1} = 4\sigma^2_i + 0.25.
\end{align}
Moreover, define $t_i$ to be the number of iterations performed on $n_i$ qubits, with $n_{1}\le n_i \le n_{m}$, and let $\ket{\psi_i}$ be the state obtained on $n_i$ qubits after $t_i$ iterations have been performed. Thus, assuming that $\H$ is acting on the appropriately sized quantum system $\ket{\psi_{i+1}}$ and $\ket{\psi_i}$ are related as follows,
\begin{align*}
    \ket{\psi_{i+1}} = \H^{t_{i+1}}\ket{\psi_i}\ket{+}.
\end{align*}
Moreover, the variance of the distribution represented by $\ket{\psi_i}\ket{+}$ is stated in Equation~\ref{eqn:variance_qubit_scaling_relation}. First, the grid on $n_i$ qubits is constant, given by $\{X_i\}$. Before applying $\H$, our distribution is simply over $\{X_i\}$, with a mean and variance given by 
\begin{align*}
    \mu_i = \sum_{x \in X_i} x P(x),\ \sigma^2_i = \sum_{x \in X_i}(x - \mu_i)^2 P(x).
\end{align*}
Upon applying $\H$, we obtain two distributions, one equal to the original distribution, and one equal to the original distribution shifted by one state to the right. The probability mass function of the new distribution would therefore be given by,
\begin{align*}
    F(x) = p P(x - 1) + (1 - p) P(x).
\end{align*}
Therefore, the mean of the new distribution would be given by,
\begin{align*}
    \mu_{i}^{\prime} = \sum_{x \in X_i}x pP(x - 1) + x (1 - p) P(x) = p(\mu_i + 1) + (1 - p)\mu_i = p + \mu_i,
\end{align*}
and the variance would be given by,
\begin{align*}
    \sigma_i^{2 \prime} &= \sum_{x \in X_i} (x - \mu_i^{\prime})^2 F(x) \\
    &= \sum_{x \in X_i} (x - \mu_i - p)^2 [p P(x - 1) + (1 - p) P(x)] \\
    &= p\sum_{x \in X_i} (x - \mu_i)^2 P(x - 1) + (1 - p)\sum_{x \in X_i}(x - \mu_i)^2P(x) \\
    &\quad + \sum_{x \in X_i}[-2xp + 2\mu_ip + p^2][pP(x - 1) + (1 - p)P(x)] \\
    &= \sigma_i^2 + p\sum_{x \in X_i}P(x) - 2p\sum_{x \in X_i}pP(x) + \sum_{x \in X_i}p^2 P(x) \\
    &= \sigma_i^2 + p(1 - p).
\end{align*}
Thus, the mean and variance on the $n_i$ qubits corresponding to the distribution obtained after $t_{i}$ iterations of $\H$ are given by 

\begin{equation}
    \mu_{i}^{\prime} = \mu_i + t_{i}p = 2\mu_{i - 1} + 0.5 + t_{i}p,
\end{equation}
\begin{equation}\label{eqn:variance_scaling_exact}
    \sigma_{i}^{2\prime} = \sigma_i^2 + t_{i}p(1 - p) = 4\sigma_{i - 1}^2 + 0.25 + t_{i}p(1 - p).
\end{equation}
Or equivalently,
\begin{equation*}
    \mu_{i}^{\prime} = 0.5\times \left( 2^{i-1}-1 \right) + \sum_{k=1}^{i} 2^{i-k}pt_k,
\end{equation*}
\begin{equation*}
    \sigma_{i}^{2\prime} = 0.25 \times \frac{4^{i-1}-1}{3} + \sum_{k=1}^{i} 4^{i-k}p(1-p)t_{k}.
\end{equation*}

It is straight forward to use this equation to both compute the effective variance obtained after performing a number of iterations at various qubit counts, and to compute the iterations required at various qubit counts to obtain a given variance. It is important to emphasize that Equation~\ref{eqn:variance_scaling_exact} implies that the qubit scaling procedure incurs \textit{no additional error} beyond that already incurred by the central limit theorem approximation to the normal distribution, as it allows for an arbitrary exact variance to be obtained. Intuitively, if you wish to obtain a distribution corresponding to $t$ iterations performed on $n_1$ qubits, if you first performed $t$ iterations on the $n_1$ qubits, and then performed some number of iterations on subsequent qubit counts, the total variance produced would exceed the desired variance. This expression describes how the number of iterations performed on the first qubit count may be reduced, and the number of removed iterations may then be translated to a number of correction iterations at higher qubit counts, allowing for the exact desired distribution to be obtained whilst still allowing for the correction iterations in the qubit scaling procedure to be made. Finally, Equation~\ref{eqn:variance_scaling_exact} shows that an arbitrary variance on $n$-qubits may be obtained while setting each $t_i$ to essentially an arbitrary constant value, so long as $t_1$ is set to give the desired variance on $n_1$ qubits. As such, it is clear that this approach implies a total number of iterations scaling polynomially in the number of qubits to obtain an arbitrary variance.

\section{Resistance to Hardware Noise} \label{appendix:noise_resistance}
The variant of the algorithm utilizing MCMR is also expected to be resistant to both bit-flip and, to a lesser extent, phase-flip errors by automatically discarding results in which such errors occur with high probability. The subsequent analysis assumes a simple noise model where errors in the execution of the algorithm are dominated by the fidelity of two-qubit gate executions. Allow $\epsilon$ to represent the two-qubit gate infidelity such that when executing any given two-qubit gate either a $X$ or a $Z$ gate (implementing a bit-flip and phase-flip error, respectively) gets subsequently randomly executed on either the control or target qubit. The probability of executing $d$ such gates without obtaining any errors is therefore given by $(1 - \epsilon)^d$.

For expository clarity, the following argument will concern itself with logical operations, such as mapping the state $\ket{j}$ to $\ket{j + 1}$, and we will not consider the Fourier basis entered by the QFT. Precisely, we will assume that the QFT and inverse QFT are applied in a noiseless channel, and that in-between these two operations the channel is subject to both phase-flip and bit-flip errors. As a result, a bit-flip error occurring in the hardware is experienced as a phase-flip error in the conceptual analytical space, and a phase-flip error on the hardware is similarly experienced as a bit-flip error in the conceptual space. This is similar to how a bit-flip error correcting code may be applied to correct a phase-flip error by applying a set of Hadamard transforms before and after the noisy channel.

Finally, we model the execution of a single iteration of $\H$ as a single quantum gate, and so an error occurring in the execution of $\H$ is modelled as one of $Z_j$ or $X_j$ being executed after $\H$ with $j$ selected uniformly at random. Note that here the notation $A_j$ means gate $A$ is executed on qubit $j$, and an identity gate is executed on all other qubits. Moreover, we assume that such errors occur with probability equal to the probability of any of the $2n$ CX gates implementing $\H$ experiencing an error. Thus, the infideltiy of the $\H$ operation is given by $1 - (1 - \epsilon)^{2n}$. In the following analysis, we will use this model to examine the impact of the first occurrence of an error after $t$ successful applications of $\H$. We will not examine the case where multiple errors occur in detail, but a similar argument holds in such cases nevertheless. We will now proceed by considering the two types of conceptual errors separately.

\paragraph{Analysis of Phase-Flip Errors} 

\begin{figure}[t]
\centering
    {
    \includegraphics[width=0.9\linewidth]{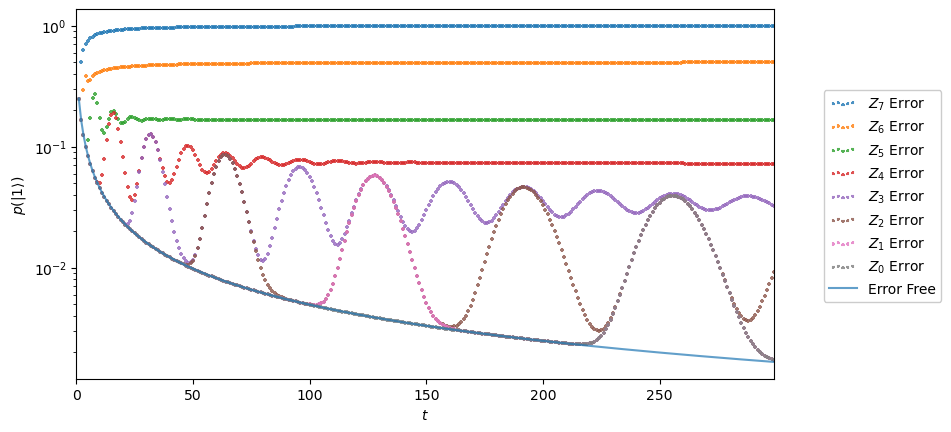}}
    \caption{\textbf{8 Qubit Demonstration of Rejection Probability as Function of Phase-Flip Errors.} In this simulated experiment, the operator $\H Z_j\H^t$ is applied to an initial state $\ket{0}$. The probability of measuring the $\ket{0}$ ancillary state in the final application of $\H$, and thus the probability of rejecting a circuit execution in which an error occurred, is shown as $p(\ket{1})$.}
    \label{fig:c_Z_error_analysis}
\end{figure}

Throughout the remainder of this document, unless we explicitly state otherwise, we allow $\{\ket{x}\}_x$ to represent the set of integer-valued standard basis vectors (i.e. $x\in\{0,1\}^n$), identically to how we previously used $\{\ket{j}\}_j$.
We now assume that we are given an initial state $\ket{\psi_i}$, apply $t$ iterations of $\H$ to obtain some state $\H^t \ket{\psi_i}$, apply an error $Z_j$ on some qubit with $0 \le j < n$, and finally apply one last iteration of $\H$, giving the final state $\H Z_j \H^t \ket{\psi_i}$. Figure~\ref{fig:c_Z_error_analysis} shows the result of this experiment, showing the probability of measuring the $\ket{1}$ ancillary state plotted for various $j$ and $t$. As can be seen in this figure, the occurrence of any error strictly increases the probability of measuring the $\ket{1}$ ancillary state, and thus of discarding the obtained error-affected result in post selection. We will now provide a general analysis of this type of error, and use the analysis to explain the plotted figure in greater detail. First, we define two new operators, $A_m = Z\otimes I \otimes ... \otimes I$, where there are $m - 1$ identity gates and one $Z$ gate, and $B_k = I \otimes ... \otimes I$ where there are $k$ identity gates, and such that $k + m = n$. We may then write,
\begin{align*}
    Z_j = B_{j} \otimes A_{n - j},
\end{align*}
where we implicitly assume that $B_0 \otimes A_{n} = A_{n}$, and that $m \ge 1$, as there must be at least one $Z$ gate in the $Z_j$ operator. First, we observe that $A_m = \diag{1\ 1\ \hdots\ -1\ -1}$, where there are $2^{m - 1}$ consecutive $+1$ elements from the start, followed by $2^{m - 1}$ elements with value $-1$. Furthermore, $B_j = \diag{1\ 1\ \hdots\ 1}$ where there are $2^j$ elements with value $+1$. By definition of the tensor product, $B_j \otimes A_{n - 1}$ thus represents a diagonal sign pattern consisting of $2^j$ repetitions of the pattern given by $A_m$ -- meaning that the period of the pattern is $2^m$. For example, in the extreme cases, $Z_0 = A_n$ represents a $2^n \times 2^n$ diagonal matrix where the first half of the diagonal terms are all $1$ and the second half of the diagonal terms are all $-1$ (thus having a period of $2^n$), and $Z_{n - 1}$ represents a $2^n \times 2^n$ diagonal matrix with the pattern $\diag{1\ -1}$ repeated $2^{n - 1}$ times. It is clear that the application of $Z_j$ to a state $\H^t \ket{\psi_i}$ will introduce a discontinuity to the wave function (beyond the negligible discontinuity already imposed by the discretization) at each pair of states $x, x+1$ for which $\text{sign}(\bra{x}Z_j\ket{x}) \neq \text{sign}(\bra{x + 1}Z_j\ket{x + 1})$. In particular, there is one such sign change in each pattern given by the diagonal of $A_{n-j}$, and so there are a maximum of $2^{j}$ possible discontinuities introduced by $Z_j$. Moreover, after $t$ iterations according to $\H$, only states $\ket{0}$ through $\ket{t-1}$ will have non-zero amplitudes, and so the number of discontinuities introduced by $Z_j$ is given by $\lfloor\frac{t}{2^{n - j}} \rfloor$. This may be made more clear by considering an example. After the application of $Z_j$ a possible state associated with the $\ket{0}$ ancillary state may look like $\ket{\psi} = \ket{x} - 2\ket{x + 1} + \ket{x + 2}$, and so $\ket{\psi} - \ket{\psi + 1} = \ket{x} - 3\ket{x + 1} + 3\ket{x + 2} - \ket{x + 3}$, as opposed to the error free state of $\ket{x} + \ket{x+1} - \ket{x+2} - \ket{x - 3}$, clearly illustrating how the sign-flip error leaves greater amplitude on the $\ket{1}$ state because of the shift discontinuous negated distribution. Indeed, SI~\ref{appendix:general_1_state_cancellation}, shows that the cancellation property of the $\ket{1}$ ancilla state only holds if the wave function is continuous. This analysis well explains the data shown in Figure~\ref{fig:c_Z_error_analysis}. Of note, as $j$ increases, as predicted by the increase in the number of discontinuities introduced, the probability of measuring the $\ket{1}$ ancillary state increases, up to a near-1 probability of discarding the execution when a $Z_7$ error occurs (in an $8$-qubit system, numbered $[0,7]$). Furthermore, when the period of the error operator's sign pattern is greater than $t$ (in particular, when it is greater than the number of states with amplitude $>>0$) the action of $Z_j$ is that of identity and so no error is incurred, hence why each of the $Z_j$ curves follows the error-free curve for some number of $t$. For example, this explains why the $Z_0$ operator follows the error-free curve until $t\approx 220$ (observing that $2^8 = 256$).

Moreover, by applying the normal approximation for the distribution produced by $t$ iterations of $\H$, it would be straight-forward to continue this analysis and obtain a closed-form expression for the probability of measuring the $\ket{1}$ state after experiencing a $Z_j$ error.

\paragraph{Analysis of Bit-Flip Errors}
\begin{figure}
\centering
    {
    \includegraphics[width=0.9\linewidth]{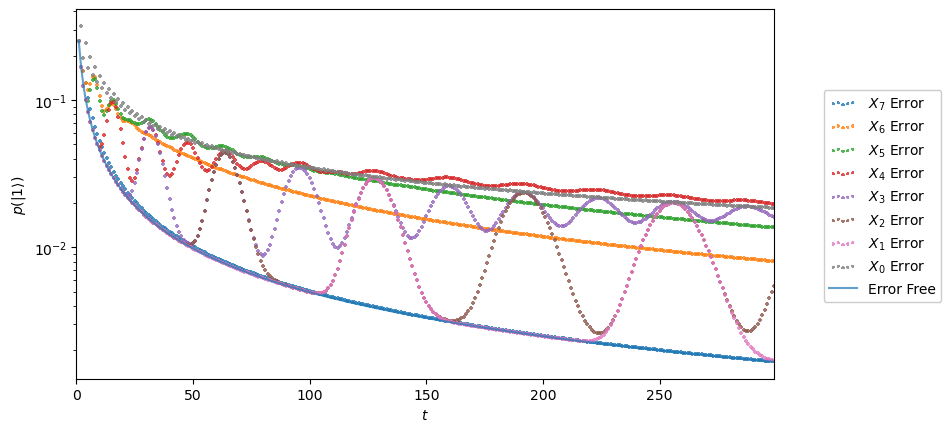}}
    \caption{\textbf{8 Qubit Demonstration of Rejection Probability as Function of Bit-Flip Errors.} In this simulated experiment, the operator $\H X_j\H^t$ is applied to an initial state $\ket{0}$. The probability of measuring the $\ket{0}$ ancillary state in the final application of $\H$, and thus the probability of rejecting a circuit execution in which an error occurred, is shown as $p(\ket{1})$.}
    \label{fig:c_X_error_analysis}
\end{figure}

The analysis for conceptual bit-flip errors follows a similar argument as that presented for the conceptual phase-flip errors, only with a couple small differences. Primarily, the discontinuities introduced by inserting an $X_j$ operator come from swapping the amplitudes of two states $\ket{x_0 \hdots x_j \hdots x_{n-1}}$ and $\ket{x_0 \hdots \overline{x}_j \hdots x_{n-1}}$. As such, as $j$ increases $X$ is applied to less-and-less significant qubits, meaning that the distance between swapped states decreases exponentially. As a result, we would expect the error introduced by the $X_j$ operator to increase as $j$ decreases, and thus for the probability of measuring the $\ket{1}$ ancillary state to increase as $j$ decreases. In the extreme, for large $n$, the distribution given by $\H^t \ket{\psi_i}$ is essentially continuous, and so swapping the amplitude on states $\ket{x}$ and $\ket{x + \delta}$ is essentially equivalent to an identity transformation, hence why the $X_7$ operator is identical to the error-free line in Figure~\ref{fig:c_X_error_analysis}. The fact that $X_j$ introduces more subtle discontinuities than $Z_j$ is also reflected in Figure~\ref{fig:c_X_error_analysis}, noting that the probability of measuring the $\ket{1}$ ancillary state (and thus of discarding a result in which an error occurred) is, in general, less than the corresponding probability for the $Z_j$ operator. We expect that an expression for the net cancellation incurred for a given $X_j$ and $t$ may be obtained by following the preceding analysis.

\section{Asymptotic analysis on the expected number of trials}\label{sec:aysmptotic_expected_trials}
In this section we analytically derive the expected number of trials to successfully load the desired distribution into a quantum register.

\subsection{Without qubit scaling}\label{sec:aysmptotic_expected_trials_no_qs}
First we look at the case where we do not perform qubit scaling in the state generation process. Suppose that we have a $n$-qubit quantum register originally in the computational basis state $\ket{x_0}$, where $x_0$ is an integer in the range of $[0, 2^n-1]$. 
Assume that $x_0 + t < 2^n$, after $t-1$ iterations of $\H$ on the quantum register, the state becomes
\begin{equation*}
    \H^{t-1} \ket{x_0} = \frac{1}{\sqrt{c_{t-1}}} \sum_{k = 0}^{t-1}\binom{t-1}{k }\ket{x_0 +k },
\end{equation*}
where
\begin{equation*}
    c_{t-1} = \sum_{k = 0}^{t-1}\binom{t-1}{k}^2
\end{equation*}
is the normalization factor. On the next iteration, the state we have before measuring the ancilla qubit is
\begin{align*}
    \frac{1}{2c_{t-1}} \sum_{k = 0}^{t-1} \left[\binom{t-1}{k}\ket{x_0 +k } + \binom{t}{k}\ket{x_0 +k +1} \right] \ket{0}\\
    + 
    \frac{1}{2c_{t-1}} \sum_{k = 0}^{t-1} \left[\binom{t-1}{k}\ket{x_0 +k } - \binom{t}{k}\ket{x_0 +k +1} \right] \ket{1} \\
    = \frac{1}{2c_{t-1}} \sum_{k = 0}^{t} \binom{t}{k}\ket{x_0 + k} \ket{0}
    + 
    \frac{1}{2c_{t-1}} \sum_{k = 0}^{t-1} \left[\binom{t-1}{k}\ket{x_0 + k} - \binom{t}{k}\ket{x_0 +k +1} \right] \ket{1}
\end{align*}
The probability of measuring $\ket{0}$ in the ancilla qubit is
\begin{align*}
    P_{t}(\ket{0}) = \frac{1}{4c_{t-1}^2}\sum_{k = 0}^{t}\binom{t}{k}^2 = \frac{c_t^2}{4c_{t-1}^2}
\end{align*}
Using the Chu–Vandermonde identity, we have
\begin{equation*}
    c_t^2 = \sum_{k = 0}^{t}\binom{t}{k}\binom{t}{k} = \binom{2t}{t}.
\end{equation*}
Therefore,
\begin{equation}
\label{eq:proba_sucess_times_c_binomial}
    P_{t}(\ket{0}) = \frac{\binom{2t}{t}}{4\binom{2t-2}{t-1}} = 1 - \frac{1}{2t}.
\end{equation}
Consequently, the probability of getting a successful state with $t$ applications of $\H$ is
\begin{equation}
\label{eq:prob_success_no_scaling}
    p_{s}(t) = \prod_{k=1}^{t}\left(1- \frac{1}{2k}\right),
\end{equation}
and the expected number of trials to get one successful run with $t$ iterations is given by the inverse of $p_{s}(t)$, i.e.
\begin{equation*}
    \mathrm{E}_t[T] = \frac{1}{p_{s}(t)} = \frac{1}{\prod_{k=1}^{t}\left(1- \frac{1}{2k}\right)}.
\end{equation*}
We take the logarithm of $\mathrm{E}_t[T]$ and expand the additive terms into Taylor series at $\infty$,
\begin{align*}
    \log(\mathrm{E}_{t}[T]) &= - \sum_{k=1}^{t}\log \left(1-\frac{1}{2k}\right)\\
    &= \sum_{k=1}^{t}\left(\frac{1}{2k} + O\left(\frac{1}{k^2}, k \to \infty\right)\right)\\
    &= \frac{1}{2}\sum_{k=1}^{t}\frac{1}{k} + O(1, t \to \infty).
\end{align*}
The last equality holds because the series of the inverse squares and higher order powers converge absolutely.
Then with the asymptotic development of harmonic series, we have
\begin{align*}
    \log(\mathrm{E}_{t}[T]) = \frac{1}{2}\log(t) + O(1, t \to \infty).
\end{align*}\\
Therefore, we have
\begin{equation}\label{eq:asymptotic_trails_binomials}
    \mathrm{E}_{t}[T] = C\sqrt{t}(1+ O(1, t \to \infty)).
\end{equation}
We will now consider an upper bound, as for any $x>-1$, $\log(1+x)\geq \frac{x}{1+x}$. In our case, $\log \left(1-\frac{1}{2k}\right) \geq \frac{-1}{2k-1}$
Thus, 
\begin{align*}
    \log(\mathrm{E}_{t}[T]) &= -\sum_{k=1}^{t}\log \left(1-\frac{1}{2k}\right)\\
    &\leq \sum_{k=1}^{t}\frac{1}{2k-1}\\
    &= 1 + \sum_{k=1}^{t-1}\frac{1}{2k+1}\\
    &\leq 1 + \sum_{k=1}^{t-1}\frac{1}{2k}
\end{align*}
Using the upper bound of the harmonic series,
\begin{align*}
    \log(\mathrm{E}_{t}[T]) \leq \frac{1}{2}\log(t-1) + \frac{3}{2}
\end{align*}
Therefore,
\begin{equation}\label{eq:upper_bound_trials_binomial}
    \mathrm{E}_{t}[T] \leq e^{\frac{3}{2}}\sqrt{t-1}
\end{equation}
Therefore the number of expected trials to get a successful state after $t$ applications of $\H$ grows sub-linearly with $t$.

\subsection{With qubit scaling}\label{sec:aysmptotic_expected_trials_qs}

First we show that the addition of a qubit in the $\ket{+}$ state on the least significant end decreases the probability of measuring the ancilla in the $\ket{1}$ state in the subsequent application of $\H$ by $1/2$, thereby increasing the algorithm's overall probability of success.
Let $a_{n,t}(x)$ be the amplitude of the $n$ qubit basis state $\ket{x}_n$ after $t$ applications of $\H$, i.e. the state in the $n$-qubit quantum register is
\begin{equation}
\label{eq:n_qubit_t_iter_state}
    \ket{\psi_t}_n = \sum_{x = 0}^{t}a_{n,t}(x)\ket{x}_n.
\end{equation}
Now we perform another application of $\H$ less the measurement, and we arrive at the state
\begin{equation*}
    \frac{1}{2} \sum_{x = 0}^{t}a_{n,t}(x)\left(\ket{x}_n + \ket{x+1}_n\right)\ket{0} + 
    \frac{1}{2} \sum_{x = 0}^{t}a_{n,t}(x)\left(\ket{x}_n - \ket{x+1}_n\right)\ket{1},
\end{equation*}
therefore the probability of measuring $\ket{1}$ in the ancilla qubit is
\begin{equation}
\label{eq:prob_1_before_adding_qubit}
    P_{t+1}(\ket{1}) = \frac{1}{4} \left\{ a_{n,t}^2(0) + a_{n,t}^2(t) + \sum_{x = 0}^{t-1}\left[a_{n,t}(x) - a_{n,t}(x+1)\right]^2 \right\}
\end{equation}
On the other hand, adding a qubit in the $\ket{+}$ state to Equation~\ref{eq:n_qubit_t_iter_state}, we have
\begin{equation*}
    \ket{\phi_t}_{n+1} = \frac{1}{\sqrt{2}} \sum_{x = 0}^{t} a_{n,t}(x) \left( \ket{2x}_{n+1} + \ket{2x+1}_{n+1} \right).
\end{equation*}
Now we perform another application of $\H$ less the measurement on $\ket{\phi_t}_{n+1}$, and we arrive at the state
\begin{align*}
    &\frac{1}{2\sqrt{2}} \sum_{x = 0}^{t}a_{n,t}(x)\left(\ket{2x}_{n+1} + 2\ket{2x+1}_{n+1} + \ket{2x+1}_{n+2}\right)\ket{0} \\
    &+ \frac{1}{2\sqrt{2}} \sum_{x = 0}^{t}a_{n,t}(x)\left(\ket{2x}_{n+1} - \ket{2x+2}_{n+1}\right)\ket{1},
\end{align*}
therefore the probability of measuring $\ket{1}$ in the ancilla qubit is
\begin{equation}
\label{eq:prob_1_after_adding_qubit}
    P'_{t+1}(\ket{1}) = \frac{1}{8} \left\{ a_{n,t}^2(0) + a_{n,t}^2(t) + \sum_{x = 0}^{t-1}\left[a_{n,t}(x) - a_{n,t}(x+1)\right]^2 \right\}.
\end{equation}
Comparing Equation~\ref{eq:prob_1_after_adding_qubit} with Equation~\ref{eq:prob_1_before_adding_qubit}, we have
\begin{equation}
\label{eq:prob_1_change_adding_qubit}
    P'_{t+1}(\ket{1}) = \frac{1}{2} P_{t+1}(\ket{1}).
\end{equation}
Therefore adding a qubit before applying $\H$ in the $\ket{+}$ state on the least significant end decreases the probability of measuring the ancilla in the $\ket{1}$ state in the subsequent application of $\H$ by $1/2$.

We now derive the expected number of trials in the context of qubit scaling. Suppose that we start with $n_1$ qubits, and gradually increase the number of qubits to $n_m$ in $m$ stages.
The qubit count increases by $1$ in each subsequent stage, and the number of applications of $\H$ is given by $t_1, t_2, \dots, t_m$, respectively.
From the analysis in Equation~\ref{appendix:efficient_scaling_qubit_addition} we know that $t_1$ is determined by the scaled variance $\hat{\sigma}$ of the target distribution, and $t_2, \dots, t_m$ are bounded by some constant number that is independent of the number of qubits in the final state. 
In other words, let $t_\text{max} = \max{t_r}, r = 2, \dots, m$, then $t_\text{max}$ is independent of $n_m$.
With this setting, the probability of getting one successful state from the process described above would be
\begin{equation*}
\label{eq:prob_success_scaling}
    p_s = \prod_{r=1}^m p_{s,r}(t_r),
\end{equation*}
where $p_{s,r}(t_r)$ is the probability of success during stage $r$.
From Equation~\ref{eq:prob_success_no_scaling} we have
\begin{equation}\label{eq:prob_success_stage_1_v2}
    p_{s,1}(t_1) = \prod_{k=1}^{t_1}\left(1 - \frac{1}{2k}\right).
\end{equation}
On the other hand, Equation~\ref{eq:prob_1_change_adding_qubit} indicates that
\begin{align}
\label{eq:prob_success_stage_r}
    p_{s,r}(t_r) >& P_{r,1}(\ket{0})^{t_r} \notag\\
    =& \left[ 1 - \frac{1}{2}P_{r-1,t_{r-1}+1}(\ket{1}) \right]^{t_r} \notag\\
    \ge& \left[ 1 - \frac{1}{2^{r-1}}P_{1,t_{1}+1}(\ket{1}) \right]^{t_r} \notag\\
    =& \left[ 1 - \frac{1}{2^{r-1}} \frac{1}{2t_1} \right]^{t_r} \notag\\
    \ge& \left[ 1 - \frac{1}{t_1} \frac{1}{2^{r}} \right]^{t_\text{max}},
\end{align}
where $P_{r,k}(\ket{0})$ and $P_{r,k}(\ket{1})$ are the probabilities of measuring $\ket{0}$ and $\ket{1}$ in the ancilla qubit after the $k$-th application of $\H$ in stage $r$.
Substituting Equation~\ref{eq:prob_success_stage_r} into Equation~\ref{eq:prob_success_scaling}, we have
\begin{align*}
    p_s >& p_{s,1}(t_1) \prod_{r=2}^m\left[ 1 - \frac{1}{t_1} \frac{1}{2^{r}} \right]^{t_\text{max}} \\
    =& p_{s,1}(t_1) \left(1 - \frac{1}{2t_1}\right)^{-t_\text{max}} \prod_{r=1}^m\left[ 1 - \frac{1}{t_1} \frac{1}{2^{r}} \right]^{t_\text{max}} \\
    =& p_{s,1}(t_1) \left[ \frac{(\frac{1}{t_1}; \frac{1}{2})_m}{1 - \frac{1}{2t_1}} \right]^{t_\text{max}} \\
    >& p_{s,1}(t_1) \left[ \frac{\left(\frac{1}{t_1}; \frac{1}{2}\right)_\infty}{1 - \frac{1}{2t_1}} \right]^{t_\text{max}},
\end{align*}
where $(a;q)_n = \prod_{k=1}^n (1 - aq^k)$ is the q-Pochhammer symbol.
And the expected number of trials to get one successful state is
\begin{equation}
\label{eq:expected_trials_scaling}
    \mathrm{E}[T] = \frac{1}{p_s} < \frac{1}{p_{s,1}(t_1)} \left[ \frac{1 - \frac{1}{2t_1}}{\left(\frac{1}{t_1}; \frac{1}{2}\right)_\infty} \right]^{t_\text{max}}.
\end{equation}
Without loss of generality, we may assume that $t_1 \ge 2$, therefore
\begin{equation*}
    p_s > p_{s,1}(t_1) \left[ \frac{\left(\frac{1}{2}; \frac{1}{2}\right)_\infty}{1 - \frac{1}{4}} \right]^{t_\text{max}} \approx 0.385^{t_\text{max}} p_{s,1}(t_1),
\end{equation*}
which is independent of $n_m$.
Consequently, the expected number of trials to get one successful state would be
\begin{equation*}
    \mathrm{E}[T] = \frac{1}{p_s} < \frac{1}{p_{s,1}(t_1)} \left[ \frac{3}{4\left(\frac{1}{2}; \frac{1}{2}\right)_\infty} \right]^{t_\text{max}} \approx \frac{2.597^{t_\text{max}}}{p_{s,1}(t_1)}.
\end{equation*}
Therefore we have proved that the expected number of trials to get one successful state is bounded by a number that is independent of the number of qubits in the final state.

Now we derive a lower bound for $\left(\frac{1}{t_1}; \frac{1}{2}\right)_\infty$ in terms of $t_1$.
To do that, we take the logarithm of the q-Pochhammer symbol, and expand each log term into Taylor series
\begin{align*}
    \log \left(\frac{1}{t_1}; \frac{1}{2}\right)_\infty
    =& \sum_{r=1}^\infty \log(1 - \frac{1}{t_1} \frac{1}{2^r}) \\
    =& - \sum_{r=1}^\infty \left[\sum_{k=1}^\infty \frac{1}{k} \left(\frac{1}{t_1 2^r} \right)^k \right] \\
    =& - \sum_{k=1}^\infty \frac{1}{kt_1^k} \left[\sum_{r=1}^\infty \left(\frac{1}{2^k} \right)^r \right] \\
    =& - \sum_{k=1}^\infty \frac{1}{kt_1^k} \frac{1}{2^k-1} \\
    =& - \frac{1}{t_1} - \sum_{k=2}^\infty \frac{1}{kt_1^k} \frac{1}{2^k-1} \\
    >& - \frac{1}{t_1} - \sum_{k=2}^\infty \frac{1}{kt_1^k} \frac{1}{2^k-2} \\
    =& - \frac{1}{t_1} - \frac{1}{2t_1}\sum_{k=1}^\infty \frac{1}{kt_1^k} \frac{1}{2^k-1} \\
    =& - \frac{1}{t_1} - \frac{1}{2t_1} \log \left(\frac{1}{t_1}; \frac{1}{2}\right)_\infty.
\end{align*}
It immediately follows that
\begin{equation*}
    \log \left(\frac{1}{t_1}; \frac{1}{2}\right)_\infty > - \frac{2}{2t_1 - 1},
\end{equation*}
or equivalently,
\begin{equation}
\label{eq:qpochhammer_bound}
    \left(\frac{1}{t_1}; \frac{1}{2}\right)_\infty > e^{-\frac{2}{2t_1 - 1}}.
\end{equation}
Substituting Equation~\ref{eq:qpochhammer_bound} back into Equation~\ref{eq:expected_trials_scaling}, we have
\begin{equation}
\label{eq:expected_trials_bound_scaling}
    \mathrm{E}[T] < \frac{1}{p_{s,1}(t_1)} \left[ \left({1 - \frac{1}{2t_1}}\right) e^{\frac{2}{2t_1 - 1}} \right]^{t_\text{max}}.
\end{equation}
We now observe that ${1 - \frac{1}{2t_1}} < 1$, and thus,
\begin{align}
    \mathrm{E}[T] < \frac{1}{p_{s,1}(t_1)}  \left[e^{\frac{2}{2t_1 - 1}}\right]^{t_\text{max}}.
\end{align}
Moreover, since we assumed that $t_1 \ge 2$, and so $e^{\frac{2}{2t_1 - 1}} < e^{\frac{2}{3}} < 2$ (i.e. $e^{\frac{2}{2t_1 - 1}}$ is maximized when $t_1$ is minimized), we get
\begin{align}
    \mathrm{E}[T] < \frac{2^{t_\text{max}}}{p_{s,1}(t_1)}.
\end{align}
From Equation~\ref{eq:prob_success_stage_1_v2}, we observe that $p_{s,1}(t_1) > (\frac{1}{2})^{t_1}$, leading to the bound
\begin{align}\label{eqn:prob_success_qubit_scaling_final_bound}
    \mathrm{E}[T] < 2^{t_\text{max} + t_1}.
\end{align}
Noting that both $t_1$ and $t_\text{max}$ are constant values independent of $n_m$, we have thus obtained a constant bound on the expected number of trials required for the procedure to succeed once. 

\section{General Proof of \texorpdfstring{$\ket{1}$}{|1>} Ancillary State Cancellation}\label{appendix:general_1_state_cancellation}
Consider some probability amplitude function function $f : \mathbb{R} \to \mathbb{R}$, such that we obtain a superposition
\begin{align*}
    \left(f(x)\ket{x} + f(x + \delta)\ket{x + \delta} \right)\ket{0} + \left(f(x)\ket{x} - f(x + \delta)\ket{x + \delta}\right)\ket{1},
\end{align*}
where $\delta$ is some small real value. The ratio of the amplitude on the $\ket{1}$ ancillary state to the total amplitude on both ancillary states may then be written as,
\begin{align}\label{eqn:general_1_state_continuous_cancellation}
    \lim_{\delta \to 0} \frac{f(x) - f(x + \delta)}{f(x) + f(x + \delta) + f(x) - f(x + \delta)} = \lim_{\delta \to 0} \frac{f(x) - f(x + \delta)}{2f(x)}
    = \lim_{\delta \to 0}\left[\frac{1}{2} - \frac{1}{2}\frac{f(x + \delta)}{f(x)}\right].
\end{align}
Suppose that $f$ is a continuous function, meaning that $\lim_{\delta \to 0}f(x + \delta) = f(x)$. Then, Equation~\ref{eqn:general_1_state_continuous_cancellation} reduces to $0$ and consequently the probability of measuring the $\ket{1}$ state is $0$ for a small shift $\delta$ when $f$ is a continuous function. Now suppose that $f$ is not continuous at point $x$, thus $\lim_{\delta \to 0}f(x + \delta) \neq f(x)$, then Equation~\ref{eqn:general_1_state_continuous_cancellation} does not necessarily simplify to zero. Indeed, it is straightforward to construct an example in which ${f(x + \delta)}/{f(x)}$ can assume an arbitrarily large or small value (by defining $f$ to be piece-wise with the left and right hand limits at $x$ being different), and thus the probability of measuring the $\ket{1}$ could be an arbitrary zero or non-zero constant.

\section{Simulation Based Approaches: Imaginary Time Evolution}\label{appendix:imaginary_time_evolution}
Another variant of the algorithm which was explored simulates $e^{\H t}$ instead of $\H^t$, where $\H\ket{x} = \ket{x - 1} + \ket{x + 1}$. A number of interesting observations were made regarding simulating $e^{\H t}$ as opposed to $\H^t$, with some distinct advantages and disadvantages noted. 

\subsection{The Need for Imaginary Time Evolution in Simulation-Based Approaches}
Originally, we considered simulating the dynamics of $\H$ in accordance with the Schrodinger equation,
\begin{align*}
    \ket{\psi(t)} = e^{-i \H t}\ket{\psi(0)}.
\end{align*}
Of course, if such a Hamiltonian $\H$ could be efficiently decomposed into some basis set (such as the set of $n$-bit Pauli strings) such that $\H = \sum_{k}t_k\H_k$ where each $\H_k$ is an element of the basis set with associated weight $t_k$, then the time evolution under $\H$ could be approximated by the Trotter-Suzuki decomposition~\cite{Trotter1959},
\begin{align*}
    e^{-i\H t} \approx \left(\prod_{k = 1}^k e^{-i \tau_k \H_k} \right)^P,
\end{align*}
where $\tau_k = \frac{t \cdot t_k}{P}$. However, in taking the Taylor expansion of $e^{-i\H t}$, we obtain,
\begin{align*}
    e^{-i\H t} = \sum_{k = 0}^{\infty} \frac{(-i)^k t^k \H^k}{k!} = \sum_{k=0, 4, ...} \frac{t^k \H^k}{k!} - i\sum_{k=1, 5, ...} \frac{t^k \H^k}{k!} - \sum_{k=2, 6, ...} \frac{t^k \H^k}{k!}
    + i\sum_{k=3, 7, ...} \frac{t^k \H^k}{k!},
\end{align*}
which corresponds to four sets of weighted sums of Binomial distributions, each of which is separated by differing phases. As a result, the states experience an undesirable interference pattern which causes a distribution that is not normal to be produced. As such, we would like to simulate the object $e^{\H t}$, as it is intuitively similar to simulating $e^{-i \H t}$ and avoids the phase complications. $e^{\H t}$ expands as
\begin{align}\label{eqn:ite_taylor_expansion}
    e^{\H t} = \sum_{k = 0}^{\infty} \frac{t^k}{k!}\H^k,
\end{align}
and as such corresponds to an infinite sum of Binomial distributions (noting that each $\H^k$ produces a particular Binomial distribution with weight $\frac{t^k}{k!}$) which is itself an approximation to a normal distribution in accordance with the central limit theorem. In understanding how $e^{\H t}$ appears to produce the same normal distribution as $\H^t$, understanding the value of $k$ in terms of $t$ for which the weights $\frac{t^k}{k!}$ are maximum may be beneficial. Whilst a more rigorous analytical argument could likely be made by substituting the factorial with an equivalent-valued Gamma function, and then computing $\frac{\partial}{\partial k}\frac{t^k}{k!} \approx \frac{\partial}{\partial k}\frac{t^k}{\Gamma(k + 1)}$, numerical simulation suggests that the coefficient is greatest when $k\approx t$. As a result, the term in Equation~\ref{eqn:ite_taylor_expansion} for which $k=t$, $\H^t$, is also likely the term which contributes the most to the transformation produced by $e^{\H t}$, providing some explanation as to why $\H^t$ appears to produce the same distributions as $e^{\H t}$. Of course, if $\H = \H^{\dagger}$, then $e^{\H t}$ will not be unitary for $t \in \mathbb{R}$. Thus, we have motivated the need for imaginary time evolution in a simulation based approach.

\subsection{Imaginary Time Evolution}
In 2017, Li and Benjamin presented a quantum-classical hybrid algorithm for the purpose of simulating the Shrodinger equation~\cite{li2017efficient}. This approach assumes that the state of the wave function at time $t$, $\ket{\phi(t)}$, may be approximated by a parameterized trial wave function $\ket{\psi(t)} \equiv | \psi(\vec{\theta}(t)) \rangle$, where $\vec{\theta}(t)$ are a set of variational time-dependant parameters. In 2019, McArdle \textit{et al.} expanded upon the aforementioned work to describe how this simulation procedure may be applied to the variational simulation of non-unitary objects such as $e^{\H t}$. In summary, the non-unitary trajectory of a given starting state is projected onto a unitary evolution through the parameters of the variational quantum circuit implementing the trial wave function.

As a result, a simulation based approach utilizing ITE for the presented distribution generating algorithm has a number of advantages and one significant challenge. The advantages are (1) the parameterized circuit captures degrees of freedom enabling shallower NISQ-friendly circuits to be obtained, (2) allows for continuous values of $t$ (as opposed to the main approach presented which requires integer values for $t$), (3) the alternating even and odd state cancellation which motivated the redefinition of the Hamiltonian as $\H\ket{x} = \ket{x} + \ket{x+1}$ is circumvented, allowing the original definition of $\H\ket{x} = \ket{x-1}+\ket{x+1}$ to be directly used, and (4) the definition of the Hamiltonian could be modified to easily enable the creation of a potentially broad class of distributions. The main challenge in using the ITE approach is that it requires the creation of a variational circuit capturing the necessary degrees of freedom in its parameters. Moreover, it is necessary to do so without introducing an exponential number of parameters (and thus number of gates) so as to remain competitive with the primary Galton simulation approaches described in this work. As such, a variational circuit should be constructed which effectively captures knowledge of the problem at hand, but how to do so in this particular instance remains a problem for future investigation.

\end{document}